%% file: elsarticle-template.tex
\documentclass[3p]{elsarticle}

\journal{Journal of Computational Physics}

\usepackage{lineno,hyperref}
\usepackage[title]{appendix}
\usepackage{amsmath}
\usepackage{dsfont}
\usepackage{array}
\usepackage{amssymb}
\usepackage{bm}
\usepackage{algorithm}
\usepackage{algpseudocode}
\usepackage{enumerate}
\usepackage{subcaption}
\usepackage{paralist}
\usepackage{multirow}
\usepackage{outlines}
\usepackage{todonotes}
\usepackage{tikz}
\usetikzlibrary{shapes}
\hypersetup{
	colorlinks,
	linkcolor={blue!50!black},
	citecolor={blue!50!black},
	urlcolor={blue!80!black},
	anchorcolor = {blue!80!black},
	filecolor = {blue!80!black},
	menucolor = {blue!80!black},
	runcolor = {blue!80!black}
}
\input{Macro.tex}

\newcommand*{\zeroCddots}{{\ddots}\kern-0.55em\raisebox{1.4ex}{\scriptsize $0$}}
\newcommand*{\oneCddots}{{\ddots}\kern-0.55em\raisebox{1.4ex}{\scriptsize $1$}}
\newcommand*{\uCddots}{{\ddots}\kern-0.55em\raisebox{1.4ex}{\scriptsize $u$}}
\newcommand*{\vCddots}{{\ddots}\kern-0.55em\raisebox{1.4ex}{\scriptsize $v$}}
\newcommand*{\wCddots}{{\ddots}\kern-0.55em\raisebox{1.4ex}{\scriptsize $w$}}

\newcommand*{\zeroCdots}{{\dots}\kern-0.8em\raisebox{0.5ex}{\scriptsize $0$}}
\newcommand*{\zeroCvdots}{{\vdots}\kern-0.0em\raisebox{0.55ex}{\scriptsize $0$}}
\AtBeginEnvironment{bmatrix}{\setlength{\arraycolsep}{4pt}}

\newcommand{\markeroneFull}{\raisebox{0.5pt}{\tikz{\node[draw,scale=0.4,circle,fill=black](){};}}}
\newcommand{\markeroneBlue}{%
  \raisebox{0.5pt}{%
    \tikz{\node[draw=blue, scale=0.4, circle](){};}%
  }%
}
\newcommand{\markeroneGreen}{%
  \raisebox{0.5pt}{%
    \tikz{\node[draw=green!50!black, scale=0.4, circle](){};}%
  }%
}
\newcommand{\markerone}{%
  \raisebox{0.5pt}{%
    \tikz{\node[draw=black, scale=0.4, circle](){};}%
  }%
}

\newcommand{\markerfour}{\raisebox{0.5pt}{\tikz{\node[draw,scale=0.4,regular polygon, regular polygon sides=4,fill=none](){};}}}
\newcommand{\markerfourBlue}{\raisebox{0.5pt}{\tikz{\node[draw=blue,scale=0.4,regular polygon, regular polygon sides=4,fill=none](){};}}}
\newcommand{\markerfourGreen}{\raisebox{0.5pt}{\tikz{\node[draw=green!50!black,scale=0.4,regular polygon, regular polygon sides=4,fill=none](){};}}}

\newcommand{\markersix}{\raisebox{0.6pt}{\tikz{\node[draw,scale=0.3,circle,fill=black!100!](){};}}}
\newcommand{\markersixBlue}{\raisebox{0.6pt}{\tikz{\node[draw,scale=0.3,circle,fill=blue!100!](){};}}}
\newcommand{\markersixGreen}{\raisebox{0.6pt}{\tikz{\node[draw=green!50!black,scale=0.3,circle,fill=green!50!black](){};}}}
\DeclareRobustCommand\full  {\tikz[baseline=-0.6ex]\draw[thick] (0,0)--(0.25,0);}
\DeclareRobustCommand\fullRed  {\tikz[baseline=-0.6ex]\draw[thick, red] (0,0)--(0.25,0);}
\DeclareRobustCommand\fullBlue  {\tikz[baseline=-0.6ex]\draw[thick, blue] (0,0)--(0.25,0);}
\DeclareRobustCommand\fullGreen  {\tikz[baseline=-0.6ex]\draw[thick, green!50!black] (0,0)--(0.25,0);}

\DeclareRobustCommand\dotted{\tikz[baseline=-0.6ex]\draw[thick,dotted] (0,0)--(0.27,0);}
\DeclareRobustCommand\dottedGreen{\tikz[baseline=-0.6ex]\draw[thick,dotted,green!50!black] (0,0)--(0.27,0);}
\DeclareRobustCommand\dottedBlue{\tikz[baseline=-0.6ex]\draw[thick,dotted,blue] (0,0)--(0.27,0);}

\DeclareRobustCommand\dashed{\tikz[baseline=-0.6ex]\draw[thick,dashed] (0,0)--(0.35,0);}
\DeclareRobustCommand\dashedGreen{\tikz[baseline=-0.6ex]\draw[thick,dashed,green!50!black] (0,0)--(0.35,0);}
\DeclareRobustCommand\dashedBlue{\tikz[baseline=-0.6ex]\draw[thick,dashed,blue] (0,0)--(0.35,0);}

\DeclareRobustCommand\chain {\tikz[baseline=-0.6ex]\draw[thick,dash dot ,green!50!black] (0,0)--(0.5,0);}
\begin{document}

\begin{frontmatter}
\title{A thermodynamically consistent and robust four-equation model for multi-phase multi-component compressible flows using ENO-type schemes including interface regularization}

\author[Stanford]{Henry Collis\corref{mycorrespondingauthor}}
\ead{hcollis@stanford.edu}
\cortext[mycorrespondingauthor]{Corresponding author}
\author[TUM]{Deniz A. Bezgin}
\ead{deniz.bezgin@tum.de}
\author[KTH,Stanford]{Shahab Mirjalili}
\ead{msey@kth.se}
\author[Stanford]{Ali Mani}
\ead{alimani@stanford.edu}
\address[Stanford]{Department of Mechanical Engineering, Stanford, CA 94305, USA}
\address[TUM]{Technical University of Munich, School of Engineering and Design, Chair of Aerodynamics and Fluid Mechanics, Boltzmannstraße 15, 85748 Garching bei München, Germany}
\address[KTH]{FLOW, Department of Engineering Mechanics, KTH Royal Institute of Technology, SE-10044 Stockholm, Sweden}

\begin{abstract}
In this work, a concise and robust computational framework is proposed to simulate compressible multi-phase multi-component flows. To handle both shocks and material interfaces, a positivity-preserving ENO-type scheme is coupled with multi-phase interface regularization terms. The positivity-preserving limiter is conservative and is applied locally for minimal degradation of the baseline ENO-type scheme. The interface regularization terms are extended from the conservative diffuse interface (CDI) model to accommodate multi-phase, multi-component flows. The ENO-type scheme is designed to be consistent with the thermodynamic equilibrium assumptions of the four-equation multi-phase model, naturally enforcing the interface equilibrium condition — preventing oscillations in pressure, velocity, and temperature around isothermal material interfaces — without requiring additional equations for volume fraction or mixture equation of state parameters, as is commonly done for the five-equation model. Additionally, non-dilute species diffusion models are extended to the multi-phase, multi-component setting. We show that this consistent framework is equally applicable for regimes ranging from single-phase to multi-phase multi-component flows. The proposed models and numerical schemes are implemented in the highly parallel Hypersonic Task based Research (HTR) Solver, and high-resolution simulations are performed using both CPUs and GPUs.

\end{abstract}


\end{frontmatter}

\section{Introduction} \label{sec:introduction}
The present work is motivated by the prevalence of multi-phase compressible flows in nature and industrial applications. These flows commonly feature interactions between shocks and material interfaces and create challenges for numerical methods. These challenges include obtaining a discrete representation of discontinuities while remaining conservative in total mass, momentum, and energy. Designing a robust scheme near discontinuities while retaining a high-resolution solution in other regions is critical to properly represent prevalent physics in many flow regimes, including turbulence. Current techniques for numerical simulation of interfaces can be broadly placed in two categories: interface tracking and interface capturing methods \cite{mirjalili2017interface}. 

\subsection{Interface tracking}

Interface tracking methods discretely represent interfaces as sharp discontinuities, where the interface is tracked using a Lagrangian representation. Some methods include Arbitrary Lagrangian-Eulerian (ALE) \cite{luo2004computation}, front-tracking \cite{glimm2003conservative,cocchi1997riemann,glimm1998three}, marker-and-cell (MAC) \cite{mckee2008mac}, and certain approaches to ghost-fluid schemes \cite{terashima2009front}. Advantages of the sharp interface include the representation of drastically different equations of state for each fluid while avoiding spurious oscillations across phase boundaries. Although interface tracking has its advantages, achieving discrete conservation is difficult and an ongoing area of research. In addition, the complexity of these methods increases when representing more complex interfacial systems, including systems with strong interfacial deformations or topological changes, high density ratios, and interactions between shocks and material interfaces. 

\subsection{Interface capturing}

Interface capturing schemes are based on an Eulerian equation to represent the location of the interface throughout time. The formulation can represent interfaces as both sharp and diffuse, where the sharp-interface methods include standard level-set/ghost-fluid schemes, \cite{mulder1992computing,fedkiw1999non,liu2003ghost,hu2004interface,chiodi2017reformulation} and geometric volume-of-fluid (VOF) \cite{debar1974fundamentals,scardovelli2003interface,aulisa2003geometrical,jofre20143,lopez2004volume,owkes2014computational,ivey2017conservative}. The sharp interface achieves similar advantages as the interface-tracking method, but the accurate geometrical representation of the interface throughout simulations without losing conservation generally becomes complex, computationally expensive, and introduces issues for scalability\cite{mirjalili2019comparison}. On the other hand, interface capturing methods including conservative level-set (CLS) \cite{olsson2005conservative}, algebraic VOF methods \cite{ubbink1999method,zhang2014refined,xie2014efficient}, and diffuse interface methods avoid the additional complexity of the sharp-interface representations by diffusing the interface over a finite number of discrete cells. The diffuse interface methods, in particular, have become a popular method due to their simplicity of implementation and load balancing, conservation guarantees, and applicability for large density ratio flows.

\subsubsection{Implicit interface capturing}
A common diffuse interface approach uses the dissipation of a numerical scheme to implicitly represent interfaces. These include total-variational-dimensioning (TVD) schemes which will keep all interfaces bounded and stable, though are generally diffusive and limited to low-order accuracy. These include flux-limiter schemes such as minmod, van Leer, and Superbee \cite{van1974towards,roe1986characteristic,van1997comparative}. High-order analogies of the TVD schemes include essentially non-oscillatory (ENO) type schemes \cite{harten1987uniformly,shu1988efficient,shu1989efficient,jiang1996efficient,fu2016family}. ENO-type schemes (WENO and TENO) use smoothness indicators to locally switch between non-dissipative high-order and dissipative low-order stencils to capture interfaces without adding substantial dissipation far from discontinuities. These techniques have traditionally been used as shock-capturing schemes for compressible flows. When ENO-type schemes are used to model immiscible interfaces, they have sometimes been referred to as interface capturing methods in analogy to their shock-capturing property \cite{johnsen2006implementation,coralic2014finite}. As described above, this terminology is also used to broadly refer to Eulerian methods for representation of interfaces, such as VOF and level-set methods. To avoid ambiguity, the term implicit interface capturing will be used in the remainder of this work to describe methods that rely on numerical dissipation, like WENO, to provide a diffuse representation of interfaces.

An advantage of many implicit interface capturing methods is their applicability to handle regions of sharp gradients due to their localized upwind-biased numerical discretization. Even so, when implicit interface capturing schemes are applied to high-Mach, high-density ratio multi-phase flow (even if combined with a phase field model \cite{Collis2022}), the small oscillations produced by the scheme can lead to simulation failure due to the development of zones which involve either negative density or internal energy. Obtaining robust solutions for high-Mach, high density ratio multi-phase flows requires enforcing positivity of mass, the squared speed-of-sound, and boundedness of phase volume fraction. These conditions are commonly achieved through the minimal and localized use of flux limiters \cite{wong2022positivity} in a manner that does not affect discrete conservation. 

A disadvantage of implicit interface capturing methods is that their dissipation can smear the material interfaces indefinitely. However, recent advancements in implicit interface capturing offer a remedy to this problem through the use of sharpening techniques from algebraic VOF methods, specifically the tangent of hyperbola for interface capturing (THINC) scheme \cite{xiao2005simple,xiao2011revisit,shyue2014eulerian,bezgin2402jax}. 

\subsubsection{Phase field methods}

In addition to implicit interface capturing methods, PDE-based diffuse interface models exist which regularize material interfaces to a finite and resolvable thickness. In particular, this work is focused on a class of diffuse interface methods known as phase field models, which includes models based on the Cahn-Hilliard \cite{cahn1958} and Allen-Cahn equations \cite{allen1979}. Recent work has proposed general implementation strategies for phase field models to provide consistent and globally conservative solutions in both incompressible and compressible flows \cite{huang2022consistent,huang2025bound}. Further advances in phase field modeling have resulted in locally conservative forms of the Allen-Cahn equation \cite{Chiu2011} known as the conservative diffuse interface (CDI) method. In recent years, significant progress has been made with the CDI model, including proving bounded volume fractions \cite{Mirjalili2020}, providing required consistency conditions for both incompressible and compressible flows \cite{Mirjalili2020,Jain2020}, and generalizing the CDI model for simulation of N-phase immiscible flows \cite{mirjalili2024} as well as on generalized curvilinear grids \cite{collis2024diffuse}. Additionally, the CDI model combined with shock-capturing schemes has been shown to be highly effective in capturing shock-interface interactions \cite{Collis2022,Jain2023}.

\subsection{Equilibrium conditions} \label{sec:Intro-EQ_cond}
In addition to a robust numerical scheme, diffuse interface methods require equilibrium conditions to define fields in the diffuse zone. The most concise model is known as the four-equation model and requires thermo-mechanical equilibrium between phases within a computational element by enforcing that one temperature, pressure, and velocity vector is shared between phases \cite{kataoka1986local,shyue1998efficient,cook2009enthalpy, Flatten2011,lemartelot2013steady, le2014towards,saurel2016general}. Other popular equilibrium conditions include a five-equation model that does not enforce thermal equilibrium \cite{allaire2002five}; the six-equation model which additionally relaxes pressure equilibrium \cite{yeom2013modified,haimovich2017numerical}; and the seven-equation model, which further relaxes momentum equilibrium between phases \cite{baer1986two}. 


We note that when modeling immiscible interfaces, diffuse interface methods artificially thicken the zone of coexisting phases solely to ease the numerical resolution requirements. Noting that physical interfaces have thicknesses on the order of a nanometer, thermo-mechanical equilibrium in the interface zone is practically instantaneous compared to the temporal resolution of numerical simulations. Therefore, using a diffuse interface model in this manner should not change the equilibrium conditions present in the sharp interface limit. However, in practice models that use additional equations compared to the four-equation model commonly sacrifice this desired thermo-mechanical equilibrium property as a trade-off to gain numerical robustness \cite{beig2015maintaining}. In some cases achieving robustness requires either sacrificing discrete conservation (of mass \cite{abgrall1996prevent,abgrall2001computations} or energy \cite{ma2017entropy})  or solving additional equations beyond the number of relaxed equilibrium conditions, e.g. equations involving the ratio of specific heats \cite{johnsen2012preventing,terashima2013consistent, nonomura2017characteristic},  which are redundant on the PDE level and lead to non-unique solutions. Aside from departing from the desired thermo-mechanical equilibrium conditions, these models incur more cost proportional to the number of additional equations which must be solved. In this sense, development of a robust and discretely conservative four-equation model without redundant PDEs addresses a physics-based need while reducing the computational cost of simulations.

\subsubsection{Interface equilibrium condition} \label{sec:Intro-IEC}

The main robustness issue with the numerical simulations of the four-equation model is formulating a scheme that satisfies the interface equilibrium condition while retaining a unique and discretely conservative solution. In the general case, the interface equilibrium condition (IEC) is defined as: \textit{for a flow with uniform pressure, temperature, and velocity, the numerical discretization should not introduce spurious oscillations in any of these fields across material interfaces}. If a scheme does not satisfy the IEC, spurious oscillations around material interfaces grow with time, potentially causing unphysical results and eventually code failure. Past work focused on satisfying the IEC with a unique and conservative four-equation model includes, \cite{fujiwara2023fully}, which proposed a conservative IEC preserving flux-splitting scheme. However, the scheme from \cite{fujiwara2023fully} is only applicable to flows involving single-phase multi-component mixtures of perfect gases without shocks. A similar extension of the method of \cite{fujiwara2023fully} to real-gas equations of state has recently been proposed \cite{terashima2025approximately}, but it can only approximately satisfy the IEC with a limited spatial order of accuracy. Other past works using the four-equation model in the compressible regime either suffered from oscillations due to lack of satisfying the IEC \cite{yi2019multicomponent}, did not achieve uniqueness or discrete conservation \cite{abgrall1996prevent,johnsen2012preventing}, used dissipative low-order schemes \cite{saurel2016general}, or resorted to dissipative spatial filters to dampen growth of the oscillations throughout time \cite{Jain2023}. 

One of the main contributions of this work is providing a computational strategy to satisfy the interface equilibrium condition near machine precision using a four-equation model. The ENO-type scheme is consistently formulated with the thermodynamic assumptions of the four-equation multi-phase model to prevent oscillations around material interfaces without additional equations, sacrificing conservation, or adding spatial filters. Additionally, the proposed IEC ENO-type scheme is applicable for general equations of state and complex mixing rules. In addition to the IEC ENO-type scheme, this work includes extensions to the CDI phase field model to multi-phase multi-component interfaces and an extension of the positivity preserving flux limiter of \cite{wong2022positivity} to the four-equation setting for robust treatment of the interactions of high-Mach shocks with high-density ratio material interfaces. The presented formulation and discretization is inclusive of an implicit interface capturing method which satisfies the IEC with the four-equation model. This can be realized by omitting the CDI regularization terms in our model. 

\subsection{Outline}

The remainder of this paper is outlined below. In Section \ref{sec:GoverningEQ} we introduce the system of equations describing multi-phase multi-component flows, the mixing rules used for interphase phases and intraphase components, and the equation of state used to closed the system. In Section \ref{sec:NumericalFramework} the spatial-temporal discretization is described, including the method used to obtain oscillation-free solutions across material interfaces using ENO-type schemes, the extension of the CDI model used to multi-phase multi-component systems, and the positivity-preserving flux-limiter for the four-equation model. Section \ref{sec:NumericalResults} presents several one-dimensional and two-dimensional cases ranging from multi-component to multi-phase multi-component flows in order to verify the application of the proposed framework. Finally, in Section \ref{sec:Conculsion}, we summarize the results and provide an outlook for future work. 

\section{Governing equations}
\label{sec:GoverningEQ}
\subsection{Physical model} \label{sec:PhysicalModel}
The multi-phase, multi-component form of the compressible Navier-Stokes equations will be studied in this work. In addition, we present a material-interface regularization model which can be added to the system to represent immiscible material interfaces. Although we can represent immiscible interfaces, the proposed model is general and can be used for modeling of miscible interfaces (similar to implicit interface capturing) by omitting the regularization terms and relying on a species diffusion model. To clearly distinguish between different phases, we introduce a notation in which the phase of a material will be indicated by $p$, where the superscript $p = 1,2$, and the components within a given phase will be indicated by the subscript $c$, where $1 \leq c \leq N$. Although we will utilize index notation for coordinates and tensor components, we do not imply index notation for $p$ and $c$. 

The fluid state at a given time $t$ can be described at position $\bm x = [x,y,z]^T$ by the vector of primitive variables $\bm W = [T, Y^c_p,u,v,w,P]^T$ or by the vector of conserved variables $\bm U = [\rho Y^c_p, \rho u, \rho v, \rho w, E]^T$. In these definitions, $Y^c_p$ is the mass of component $c$ of phase $p$ per total mass, known as the mixture mass fraction, $\bm u = [u,v,w]^T$ is the velocity vector, $P$ is the pressure, $T$ is the temperature, $\rho$ is the mixture density, and $E$ is the total energy per unit volume defined as $E = \rho e + \frac{1}{2}\rho \bm u \cdot \bm u$ where $e$ is the internal energy per unit volume. 
Using these variables, the multi-phase multi-component Navier-Stokes equations with interface regularization terms can be compactly written in differential form in terms of the conserved variables $\bm U$,

\begin{align} \label{eq:GoverningEquations}
    \frac{\partial \bm U}{\partial t} + \frac{\partial \left [\mathcal{F}(\bm U)+\mathcal{F_{\nu}}(\bm U)+\mathcal{F_{DI}}(\bm U)\right ]}{\partial x} + \frac{\partial \left [\mathcal{G}(\bm U)+\mathcal{G_{\nu}}(\bm U)+\mathcal{G_{DI}}(\bm U)\right ]}{\partial y} + \frac{\partial \left[\mathcal{H}(\bm U) + \mathcal{H_{\nu}}(\bm U)+\mathcal{H_{DI}}(\bm U)\right ]}{\partial z} = 0,
\end{align}
for $p=1,2$ and $1\le c\le N$. The convective fluxes $\mathcal{F}$, $\mathcal{G}$, $\mathcal{H}$ are defined as,

\begin{align} \label{eq:ConvectiveTerms}
    \mathcal{F}(\bm U) = \begin{bmatrix}
    \rho Y^c_pu \\
    \rho uu + P \\ \rho uv \\ \rho uw  \\ u(E+P)
    \end{bmatrix} \quad 
    \mathcal{G}(\bm U) = \begin{bmatrix}
    \rho Y^c_pv \\
    \rho uv \\ \rho vv + P \\ \rho vw  \\ v(E+P)
    \end{bmatrix} \quad
    \mathcal{H}(\bm U) = \begin{bmatrix}
    \rho Y^c_pu \\
    \rho uw \\ \rho vw \\ \rho ww + P \\ w(E+P)
    \end{bmatrix}, 
\end{align}
and the viscous fluxes, $\mathcal{F_{\nu}}$, $\mathcal{G_{\nu}}$, $\mathcal{H_{\nu}}$ can be expressed as,

\begin{align} \label{eq:DiffusionTerms}
    \mathcal{F_{\nu}}(\bm U) = \begin{bmatrix}
    J_{p,x}^c \\
    -\tau_{11} \\ -\tau_{12} \\ -\tau_{13}  \\ \Sigma_iu_i\tau_{1i} - q_1
    \end{bmatrix} \quad 
    \mathcal{G_{\nu}}(\bm U) = \begin{bmatrix}
    J_{p,y}^c \\
    -\tau_{21} \\ -\tau_{22} \\ -\tau_{23}  \\ \Sigma_iu_i\tau_{2i} - q_2
    \end{bmatrix} \quad
    \mathcal{H_{\nu}}(\bm U) = \begin{bmatrix}
    J_{p,z}^c \\
    -\tau_{31} \\ -\tau_{32} \\ -\tau_{33}  \\ \Sigma_iu_i\tau_{3i} - q_3
    \end{bmatrix}, 
\end{align}
where $\tau_{ij}$ is the viscous stress tensor,

\begin{equation}
    \tau_{ij} = \mu\left[\frac{\partial u_i}{\partial x_j} + \frac{\partial u_j}{\partial x_i} - 2/3\frac{\partial u_k}{\partial x_k}\delta_{ij}\right],
\end{equation} 
with $\mu$ as the dynamic viscosity, and $\delta_{ij}$ as the Kronecker delta. Additionally, the intraphase species mass diffusion vector can be derived for non-dilute species diffusion from the Stefan-Maxwell diffusion model (with certain equilibrium assumptions) as a Fickian diffusion term and a mass corrector \cite{coffee1981transport}. This well-known model for non-dilute species diffusion can be extended to the multi-phase multi-component context using a confined scalar argument. The resulting model can be written as,
\begin{align}
  \label{eq:SpeciesMassDiffusion}
    J_{p,i}^c & = - \rho Y_p^c \left[D^c_p \frac{\partial }{\partial \bm x_i} \left(\ln \left(\frac{X_p^c}{X_p}\right)\right) - \sum_j \left(\frac{Y_p^j}{Y_p}\right) D^j_p \frac{\partial }{\partial \bm x_i}\left(\ln \left(\frac{X_p^j}{X_p}\right)\right)\right] 
    \\
    & = - \rho Y_p^c \left[D^c_p\frac{X_p}{X_p^c} \frac{\partial }{\partial \bm x_i}  \left(\frac{X_p^c}{X_p}\right) - \sum_j \left(\frac{Y_p^j}{Y_p}\frac{X_p}{X_p^j}\right) D^j_p \frac{\partial }{\partial \bm x_i} \left(\frac{X_p^j}{X_p}\right)\right] \\
    & = - \rho  \left[ D^c_pY_p\frac{W_p^c}{W_p} \frac{\partial }{\partial \bm x_i}  \left(\frac{X_p^c}{X_p}\right) - Y_p^c\sum_j \left(\frac{W_p^j}{W_p}\right) D^j_p \frac{\partial }{\partial \bm x_i} \left(\frac{X_p^j}{X_p}\right)\right]
\end{align} 
where $D^c_p$ is the mass diffusivity of component $c$ within phase $p$, $X_p^c$ is the mixture molar fraction of component $c$ of phase $p$, $X_p$ is the mixture molar fraction of phase $p$, $Y_p$ is the mixture mass fraction of phase $p$, $W_p^c$ is the molecular weight of component $c$, and $W_p$ is the molecular weight of phase $p$. Formulas describing the computation of $W_p$, $Y_p$, and $X_p$ are in \ref{sec:Mixing rules}. 
One key difference between this formulation and the common non-dilute Stefan-Maxwell model is the incorporation of the denominators ($X_p$ and $Y_p$) in the gradient terms. This restricts the volume of influence of species diffusion to be within each phase and prevents unphysical leakage of components across phases. This formulation can be derived using the consistent transport models for confined scalars introduced in previous work \cite{mirjalili2022computational,mirjalili2022consistent}. Any physical exchange across phases (e.g. phase change) would require additional models to explicitly represent mass-transfer between phases. Lastly, the heat flux $q_i$ is defined as,

\begin{equation}
    q_i = -\lambda \frac{\partial T}{\partial x_i} + \sum_p\sum_c J_{p,i}^c \frac{Y_p^c}{Y_p}h_p^c,
\end{equation}
where $\lambda$ is the heat conductivity of the mixture and $h_p^c$ is the specific enthalpy of component $c$ of phase $p$. The final terms in Eq. \ref{eq:GoverningEquations}, $\mathcal{F_{DI}}$, $\mathcal{G_{DI}}$, $\mathcal{H_{DI}}$, are the interface regularization terms (i.e. the diffuse interface terms) and are described in Section \ref{sec:DiffuseInterfaceTerms}.

\subsection{Mixture rules} \label{sec:MixtureRules}

\subsubsection{Equilibrium assumptions}
In this work, thermal and mechanical equilibrium is assumed between phases within a computational element by enforcing that one temperature, pressure, and velocity vector is shared between phases. In the literature, models which enforce thermo-mechanical equilibrium are often known as the four-equation multi-phase model \cite{kataoka1986local,shyue1998efficient,cook2009enthalpy, Flatten2011,lemartelot2013steady, le2014towards,saurel2016general}. One of the main contributions of this work is providing a  computational strategy to satisfy the discrete interface equilibrium condition near machine precision  with the four-equation model. The proposed scheme is presented in Section \ref{sec:spatialDiscretization}. 

The following sections overview the mixing rules required to obtain the equation of state closure for Eq. \ref{eq:GoverningEquations}. Section \ref{sec:InterphaseMixing} starts by defining the mixing rules describing multi-phase mixtures and Section \ref{sec:IntraphaseMixing} extends this to a general multi-phase multi-component model.  

\subsubsection{Interphase mixing rules} \label{sec:InterphaseMixing}
In this work, multi-phase mixtures are assumed to be immiscible. To describe immiscible phases, separate phases are assumed to occupy their own individual volumes and share a common pressure (Amagat's law). The mixing rules are summarized as,
\begin{equation}
    T_p = T \quad \forall p, \quad
    P_p = P \quad \forall p, \quad
    v = \sum_p Y_p v_p, \quad
    e = \sum_p Y_p e_p
\end{equation}
where $T_p$, $P_p$, $e_p$, and $v_p$ are the temperature, pressure, internal energy, and specific volume for phase $p$. These thermodynamic variables can be obtained from the equation of states for each phase $p$. These mixing rules allow for each phase $p$ to be governed by different equation of states as well as by unique intraphase mixing rules to describe multi-component mixtures.

Other quantities that are treated with interphase mixing rules include the mixture viscosity $\mu$ and mixture thermal conductivity $\lambda$ as,
\begin{equation}
    \mu = \sum_p \phi_p\mu_p, \quad
    \lambda = \sum_p \phi_p\lambda_p
\end{equation}
where $\mu_p$ and $\lambda_p$ are the phasic viscosity and thermal conductivity respectively, and $\phi_p$ is the phase volume fraction. In this work $\lambda_p$ and $\mu_p$ are assumed constant in the liquid phase. For the gaseous phase, $\lambda_{g}$ is defined as,
\begin{equation}
    \lambda_{g} = \frac{\mu_{g}}{\text{Pr}}\sum_c(Y_g^c/Y_g)C{_P}_g^c
\end{equation}
and $\mu_g$ is defined using Wilke's rule \cite{wilke1950viscosity} as,
\begin{equation}
    \mu_g = \sum_i\frac{(Y_g^i/Y_g)\mu_g^i}{\sum_j G_{ij}W_c/W_j(Y_g^j/Y_g)}
\end{equation}
where,
\begin{equation}
    G_{ij} = \frac{1}{\sqrt{8}}\left(1+\frac{W_i}{W_j}\right)^{-1/2}\left[1 + \left(\frac{\mu_g^i}{\mu_g^j}\right)^{-1/2}\left(\frac{W_i}{W_j}\right)^{1/4}\right]^{2}
\end{equation}
and the Prandtl number, Pr, the dynamic viscosity of species $\mu_g^i$, and the specific heats, $C{_P}_g^c$, are assumed constant. Although not explored in this work, the formulation does not restrict these quantities from being extended to more advanced definitions, including defining the components with a non-linear dependence on temperature.  

\subsubsection{Intraphase mixing rules} \label{sec:IntraphaseMixing}
Classical intraphase mixing rules include assuming the individual components within a phase occupy their own volume (Amagat's law) or assuming components share volumes (Dalton's law). Traditionally, mixtures of gas components are assumed to follow Dalton's law, with each individual component contributing a partial pressure towards the mixture pressure \cite{CHIAPOLINO201731}. For a real gas equation of state (e.g. including the compressiblity factor), Dalton's law and Amagat's law will give different results. However, if the gas components are modeled as ideal gases, both representations obtain equivalent thermodynamic states \cite{CHIAPOLINO201731}. 

The governing equations in Section \ref{sec:PhysicalModel} are written to allow either intraphase mixing rule to be applied. In this work, though the gas components will follow the ideal gas law, both mixing rules will be summarized to show the generality of the four-equation model and provide a background for extensions to more complex equations of state. 

The ideal mixing rules associated with assuming all intraphase components occupy the same volume (Dalton's law) are summarized as,
\begin{equation}
    T_p = T^c \quad \forall c, \quad
    P_p = \sum_c P^c_{partial}, \quad
    v_p = v^c_p \quad \forall c, \quad 
    e_p = \sum_c (Y^c_p/Y_p) e^c_p
\end{equation}
where (assuming the ideal gas law) the partial pressure of component $c$ is defined as,

\begin{equation}
    P^c_{partial} = \rho Y_c^p R T /W^c
\end{equation}
where $R$ is the universal gas constant and $W^c$ is the molecular weight of component $c$. Additionally, the ideal mixing rules associated with assuming all components occupy the individual volumes (Amagat's law) are summarized as,
\begin{equation}
\begin{split}
    T_p = T^c \quad \forall c, \quad
    P_p = P^c \quad \forall c, \quad
    v_p = \sum_c (Y_p^c/Y_p) v^c_p, \quad
    e_p = \sum_c (Y_p^c/Y_p) e^c_p.
\end{split}
\end{equation}
Lastly, the mass diffusivity of component $c$ within phase $p$ is defined as,

\begin{equation}
    D_p^c = \frac{\mu_p}{\rho_p \text{Sc}_p^c}
\end{equation}
where, in this work, a constant Schmidt number, $\text{Sc}_p^c$, is assumed for each component within a phase.

\subsection{Noble Abel stiffened gas EOS}

To close the system illustrated in Eq. \ref{eq:GoverningEquations} the Nobel-Abel equation of state (NASG EOS) \cite{le2016noble} is used for all components in the multi-phase multi-component mixture. For a pure component, the general NASG EOS assuming constant heat capacity reads,

\begin{equation} \label{eq:NASG_EOS}
\begin{split}
&v_p^c(P,T) = \frac{(C{_P}^c_{p} - C{_v}^c_{p})T}{P + P{_\infty}_p^c} + b_p^c \\
&e_p^c(P,T) = \frac{P + \gamma_p^c P{_\infty}_p^c}{P + P{_\infty}_p^c}C{_v}^c_{p}T + q_p^c \\
&h_p^c(P,T) = C{_P}^c_{p} T + b_p^c P + q_p^c \\
\end{split}
\end{equation}
where for component $c$ in phase $p$, $C{_P}^c_{p}$ is the heat capacity at constant pressure, $C{_v}^c_{p}$ is the heat capacity at constant volume, $\gamma_p^c$ is the heat capacity ratio, $b_p^c$ is the co-volume of the component molecules, $P{_\infty}_p^c$ is the stiffened pressure to model the attractive forces between molecules in a material, and $q_p^c$ is a reference energy. If $b_p^c$ and $P{_\infty}_p^c$ are taken as zero (as is done for components in the gas phase) the NASG EOS reduces to the ideal gas equation of state. 

During a simulation, the internal energy and volume (or density) of the mixture can be determined from the conserved vector $\bm U$ and the mixing rules in Section \ref{sec:MixtureRules}. To obtain expressions for the pressure and temperature of the mixture in terms of internal energy and volume, the mixing rules discussed in Section \ref{sec:MixtureRules} can be combined with the NASG EOS. For a system with multiple components in both phases, an iterative procedure is required to find the pressure and temperature that satisfy the equilibrium conditions\cite{cook2009enthalpy}. In this work, only one component will be modeled in the liquid state, and the gaseous state will be composed of ideal gas components. Given these simplifications, we can use Amagat's law for both intraphase and interphase mixing and the NASG EOS to obtain closed expressions for the mixture pressure. An expression for the mixture pressure can be found by enforcing the equality between the two definitions for mixture temperature found from Eq. \ref{eq:NASG_EOS},

\begin{align}
    &T(e, P, Y_p^c) = (e-q)\left(\sum_p\sum_c\frac{Y_p^cC{_v}_p^c(P + \gamma_p^cP{_\infty}_{p}^{c})}{P + P{_\infty}_{p}^{c}}\right)^{-1} \label{eq:NASG_T1}\\
    &T(v, P, Y_p^c) = (v-b)\left(\sum_p\sum_c\frac{Y_p^cC{_v}_p^c(\gamma_p^c-1)}{P + P{_\infty}_{p}^{c}}\right)^{-1}.\label{eq:NASG_T2}
\end{align}
Solving for a common pressure between Eq. \ref{eq:NASG_T1} and \ref{eq:NASG_T2} leads to
\begin{equation} \label{eq:NASG_P}
    P = \frac{a_2 + \sqrt{a_2^2 + 4a_1a_3}}{2a_1}
\end{equation}
with
\begin{equation}
\begin{split}
    & a_1 = C_v \\
    & a_2 = \left(\frac{e - q}{v - b}\right)(C_P - C_v) - P{_\infty}C_v - P{_\infty}Y_1(C{_P}_1 - C{_v}_1) \\
    & a_3 = \left(\frac{e - q}{v - b}\right)P{_\infty}[C_P - C_v - Y_1(C{_P}_1 - C{_v}_1)]
\end{split}
\end{equation}
and
\begin{equation}
    C_v = \sum_p\sum_cY_p^cC{_v}_p^c, \quad C_P = \sum_p\sum_cY_p^cC{_P}_p^c, \quad q = \sum_p\sum_cY_p^cq_p^c, \quad b= \sum_p\sum_cY_p^cb_p^c
\end{equation}
where for this work $p=1$ is defined as the liquid phase and $P{_\infty} \equiv P{_\infty}_{p=1}^{c=1}$ as there is only one component in the liquid. Once pressure is determined using Eq. \ref{eq:NASG_P}, either Eq. \ref{eq:NASG_T1} or Eq. \ref{eq:NASG_T2} can be used to obtain the mixture temperature. For mixtures with multiple components in each phase or specific heats dependent on temperature, iterative solvers can be used to determine the equilibrium state \cite{cook2009enthalpy}.

Unless otherwise specified, the NASG EOS parameters used for all materials throughout this work are included in Table \ref{tab:NASGMaterialParamters}.

\begin{table}[hbt!]
    \centering
    \begin{tabular}{l c c c c c}
    \hline
     Material & $C_P$ [$\text{J}$ $ \text{kg}^{-1}\text{K}^{-1}$] & $\gamma$ [-] & q [$\text{J}$ $ \text{kg}^{-1}$] & b [$\text{kg}^{-1}$] & $P_{\infty}$ [Pa] 
    \\[1mm]
    \hline
    Water (Liquid) & $4.185\times 10^3$ & $1.0123$ & $-1.143\times10^6$ & $9.203\times 10^{-4}$ & $1.835 \times 10^8$
    \\[1mm]
    Air & $1.011\times10^3$ & 1.4 & 0 & 0 & 0
    \\[1mm]
    Helium & $5.091\times10^3$ & 1.66 & 0 & 0 & 0
    \\[1mm]
    SF6 & $0.661\times10^3$ & 1.093 & 0 & 0 & 0
    \\[1mm]
    \hline
    \\
    \end{tabular}
    \caption{NASG Parameters used in this work}
    \label{tab:NASGMaterialParamters}
\end{table}

\section{Numerical framework}
\label{sec:NumericalFramework}
\subsection{Spatial discretization} \label{sec:spatialDiscretization}
In this work, the integral form of the partial differential equations in Eq. \ref{eq:GoverningEquations} will be solved using the finite-volume method. The differential form of Eq. \ref{eq:GoverningEquations} is assumed to admit smooth solutions for which partial derivatives exist. A general cuboid cell $(i,j,k)$ has the spatial width defined by $\Delta x$, $\Delta y$, and $\Delta z$ and a cell volume defined as $\Delta x\Delta y\Delta z$. In this work, isotropic Cartesian grids ($\Delta x = \Delta y = \Delta z$) will be used to spatially discretize the system. In the finite-volume formulation the cell averaged values of the conserved variables can be obtained as,

\begin{equation}
    \Bar{\bm U}_{i,j,k} = \frac{1}{V}\int_{x_{i-\frac{1}{2},j,k}}^{x_{i+\frac{1}{2},j,k}}\int_{y_{i,j-\frac{1}{2},k}}^{y_{i,j+\frac{1}{2},k}}\int_{z_{i,j,k-\frac{1}{2}}}^{z_{i,j,k+\frac{1}{2}}}\bm U dxdydz.
\end{equation}

Applying the volume integration to Eq. \ref{eq:GoverningEquations} leads to a set of ordinary differential equations,

\begin{equation} \label{eq:DiscreteGoverningEquations}
\begin{split}
    \frac{d}{dt}\Bar{\bm U}_{i,j,k} = & -\frac{1}{\Delta x}\left[\mathcal{F}_{i+\frac{1}{2},j,k} - \mathcal{F}_{i-\frac{1}{2},j,k}\right] - \frac{1}{\Delta y}\left[\mathcal{ G}_{i,j+\frac{1}{2},k} - \mathcal{G}_{i.j-\frac{1}{2},k}\right] - \frac{1}{\Delta z}\left[\mathcal{ H}_{i,j,k+\frac{1}{2}} - \mathcal{H}_{i,j,k-\frac{1}{2}}\right] \\
    & -\frac{1}{\Delta x}\left[\mathcal{F{_\nu}}_{i+\frac{1}{2},j,k} - \mathcal{F{_\nu}}_{i-\frac{1}{2},j,k}\right] - \frac{1}{\Delta y}\left[ \mathcal{G{_\nu}}_{i,j+\frac{1}{2},k} - \mathcal{G{_\nu}}_{i.j-\frac{1}{2},k}\right] - \frac{1}{\Delta z}\left[\mathcal{H{_\nu}}_{i,j,k+\frac{1}{2}} - \mathcal{H{_\nu}}_{i,j,k-\frac{1}{2}}\right] \\
    & -\frac{1}{\Delta x}\left[\mathcal{F{_{DI}}}_{i+\frac{1}{2},j,k} - \mathcal{F{_{DI}}}_{i-\frac{1}{2},j,k}\right] - \frac{1}{\Delta y}\left[\mathcal{G{_{DI}}}_{i,j+\frac{1}{2},k} - \mathcal{G{_{DI}}}_{i.j-\frac{1}{2},k}\right] - \frac{1}{\Delta z}\left[\mathcal{ H{_{DI}}}_{i,j,k+\frac{1}{2}} - \mathcal{H{_{DI}}}_{i,j,k-\frac{1}{2}}\right]
\end{split}
\end{equation}
where, $\mathcal{F}_{i+1/2, j, k}$, $\mathcal{F_\nu}_{i+1/2, j, k}$, and $\mathcal{F_{DI}}_{i+1/2, j, k}$ are the averaged cell face fluxes on the $x_{i+1/2}$ face for the convective, diffusive, and regularization terms respectively. The definitions of the fluxes on the $y-$ and $z-$ faces are analogous. For multi-dimensional problems Eq. \ref{eq:DiscreteGoverningEquations} is solved using a dimension-by-dimension approach where each dimension is evaluated separately. Additionally, in this work all cell-face averaged variables are approximated using a one-point quadrature rule. Higher-order representations are possible and have been explored in other works \cite{coralic2014finite}. To calculate the convective flux on the cell-faces the Godunov approach \cite{toro2013riemann} is used by solving a Riemann problem on a cell face. The Riemann problem is defined by fluid states on the left and right sides of a cell face, which are obtained by numerical interpolations. Across discontinuities, including shocks and material interfaces, the method of interpolation can be critical to obtaining a robust and oscillation-free solution. A brief review of previous approaches to limiting oscillations across discontinues and interfaces was presented in Section \ref{sec:introduction}. 

In this work, an ENO-type (WENO-type, or TENO-type) interpolation scheme is used in combination with a HLLC Riemann solver. A key challenge when applying this methodology to material interfaces is formulating the interpolations to satisfy the IEC. Details on the variables which are reconstructed by the ENO-type scheme while satisfying the IEC can be found in Section \ref{sec:IEC} and details on the HLLC form is in \ref{sec:HLLC}. The viscous fluxes are interpolated to faces using a standard second-order central scheme. The interface regularization fluxes are interpolated to faces using a second-order kinetic energy and entropy preserving (KEEP) central interpolation that satisfies the IEC \cite{Jain2022KEEP}. More detail on the regularization terms and the KEEP interpolation is in Section \ref{sec:DiffuseInterfaceTerms}. 
  
\subsubsection{Avoiding spurious oscillations} \label{sec:AvoidingSpuriousOscillations}
High-order interpolation of the fluid state directly to the cell-faces can result in spurious oscillations around interfaces due to the interactions of discontinuities in dependent fields. Additionally, the non-linearity of ENO-type schemes can lead to different numerical stencils for each fluid field. As these fields are dependent on one another in physical space, the inconsistency between the stencils (and the stencil's numerical dissipation) can result in spurious oscillations. Instead, projecting the fields to characteristic space before interpolation decouples the fluid fluids into independent characteristic variables and avoids the issues associated with using ENO-type schemes \cite{harten1987uniformly,shu1988efficient,shu1989efficient,shu1992high}. 

Although interpolating the characteristic variables reduces spurious oscillations, it does not cause the system to inherently satisfy the interface equilibrium condition and will still result in oscillations around material interfaces. As discussed in Sections \ref{sec:Intro-EQ_cond} and \ref{sec:Intro-IEC} past works have presented strategies for satisfying the IEC. However, these strategies either require loss of conservation, add redundant equations, or restrict mixing rule and thermodynamic equilibrium assumptions.


\subsubsection{Satisfying the interface equilibrium condition} \label{sec:IEC}

Notably, in this work, the IEC is satisfied to near-machine precision using the four-equation model without redundant equations. The idea is based on the use of the primitive vector $\bm W = [T,Y^c_p,u,v,w,P]^T$, as the basis for building the interpolations to cell faces. An important clarification must be made that this set of primitive variables includes temperature instead of density (unlike what is done for the volume-fraction-based five-equation model \cite{coralic2014finite}). Interpolating (including with a nonlinear ENO-type scheme) directly based on $P$, $T$, and $\bm u$ enforces the interface equilibrium condition numerically, since a well-formulated numerical scheme will keep a constant field constant after interpolation. Replacement of $T$ with $\rho$, as done in other studies \cite{coralic2014finite}, will no longer maintain a constant temperature after interpolation (since $\rho$ contains a jump across the interface) and results in oscillations of $P$, $T$, and $\bm u$ (shown in Table \ref{tab:DropAdvectionNoIECResults}). Instead, using the primitive vector, $\bm W = [T,Y^c_p,u,v,w,P]^T$, removes oscillations around isothermal material interfaces and discretely satisfies the IEC. 

Though the strategy outlined above will satisfy the IEC, as discussed in Section \ref{sec:AvoidingSpuriousOscillations}, it is advantageous (especially around shocks) to decompose the fluid-state into characteristic variables in order to avoid additional spurious oscillations from the non-linearity of the ENO-type schemes. In this work, the proposed primitive basis of $\bm W = [T,Y^c_p,u,v,w,P]^T$ (not including one $Y^c_p$) is used to define characteristic variables. The characteristic matrices used for the projection of $\bm W = [T,Y^c_p,u,v,w,P]^T$ into characteristic space are included in \ref{sec:CharacteristicForm}. Additionally, using an interpolation based on characteristic variables with a non-linear ENO-type scheme can be analytically shown to satisfy the IEC ( \ref{sec:IECCharacteristicAnalysis}). A numerical example showcasing the successful application of this IEC ENO-type scheme to solve an inviscid multi-phase droplet advection with oscillations in pressure, velocity, or temperature, near machine precision is in Section \ref{sec:DropAdvection}.

\subsubsection{Godunov Approach} \label{sec:GodunovApproach}
The resulting numerical strategy for obtaining the flux on cell faces is summarized below.

\begin{enumerate}
    \item Project $\bm W$ into characteristic space to obtain $\Tilde{\bm W}$
    \begin{enumerate}
        \item Define an average of $\bm U$ on a face: $\bm U_{i+\frac{1}{2}}$ (using either arithmetic or Roe average)
        \item Define the left eigenvector, $\bm S^{-1}$ from the averaged state of $\bm U_{i+\frac{1}{2}}$
        \item $\Tilde{\bm W}$ =  $\bm S^{-1}\bm W$
    \end{enumerate}
    \item Obtain left and right states of $\Tilde{\bm W}$ on the cell face using ENO-type scheme 
    \item Project $\Tilde{\bm W}^L_{i+\frac{1}{2}}$ and $\Tilde{\bm W}^R_{i+\frac{1}{2}}$ to physical space to obtain $\bm W^L_{i+\frac{1}{2}}$ and $\bm W^R_{i+\frac{1}{2}}$
    \begin{enumerate}
        \item Define the right eigenvector, $\bm S$ from the averaged state of $\bm U_{i+\frac{1}{2}}$
        \item $\bm W^L_{i+\frac{1}{2}}$ =  $\bm S\Tilde{\bm W}^L_{i+\frac{1}{2}}$, $\bm W^R_{i+\frac{1}{2}}$ =  $\bm S\Tilde{\bm W}^R_{i+\frac{1}{2}}$
    \end{enumerate}
    \item Obtain the left and right states for density using the EOS
    \item Find the left and right states of the conserved variables from primitives
    \item Find the cell-face flux using a HLLC Riemann solver
    \begin{enumerate}
        \item ${\mathcal{F}}_{i+\frac{1}{2}}$ =  \textbf{HLLC}$(\bm U^L_{i+\frac{1}{2}},\bm U^R_{i+\frac{1}{2}})$
    \end{enumerate}
\end{enumerate}

The approach outlined above will result in high-resolution and low-dissipation capturing of material interfaces and shocks. However, unlike shocks, material interfaces are not self-sharpening. Though high-order ENO-type interpolations can be used across material interfaces as an implicit interface capturing scheme, the numerical dissipation from the upwind-biased interpolations will cause mixing of immiscible phases throughout time. To enforce immiscibility between phases, interface regularization terms can be added to the system to sharpen multi-phase interfaces while allowing intraphase species diffusion.

\subsection{Interface regularization} \label{sec:DiffuseInterfaceTerms}
To counteract the intrinsic numerical diffusion caused by both the ENO-type interpolation and the approximate Riemann solver \cite{toro2013riemann} from the Godunov approach described in Section \ref{sec:GodunovApproach}, material interface regularization terms can be added to enforce a finite thickness interface throughout a simulation. 

\begin{figure}[hbt!]
\begin{center}
\includegraphics[width=0.8\textwidth]{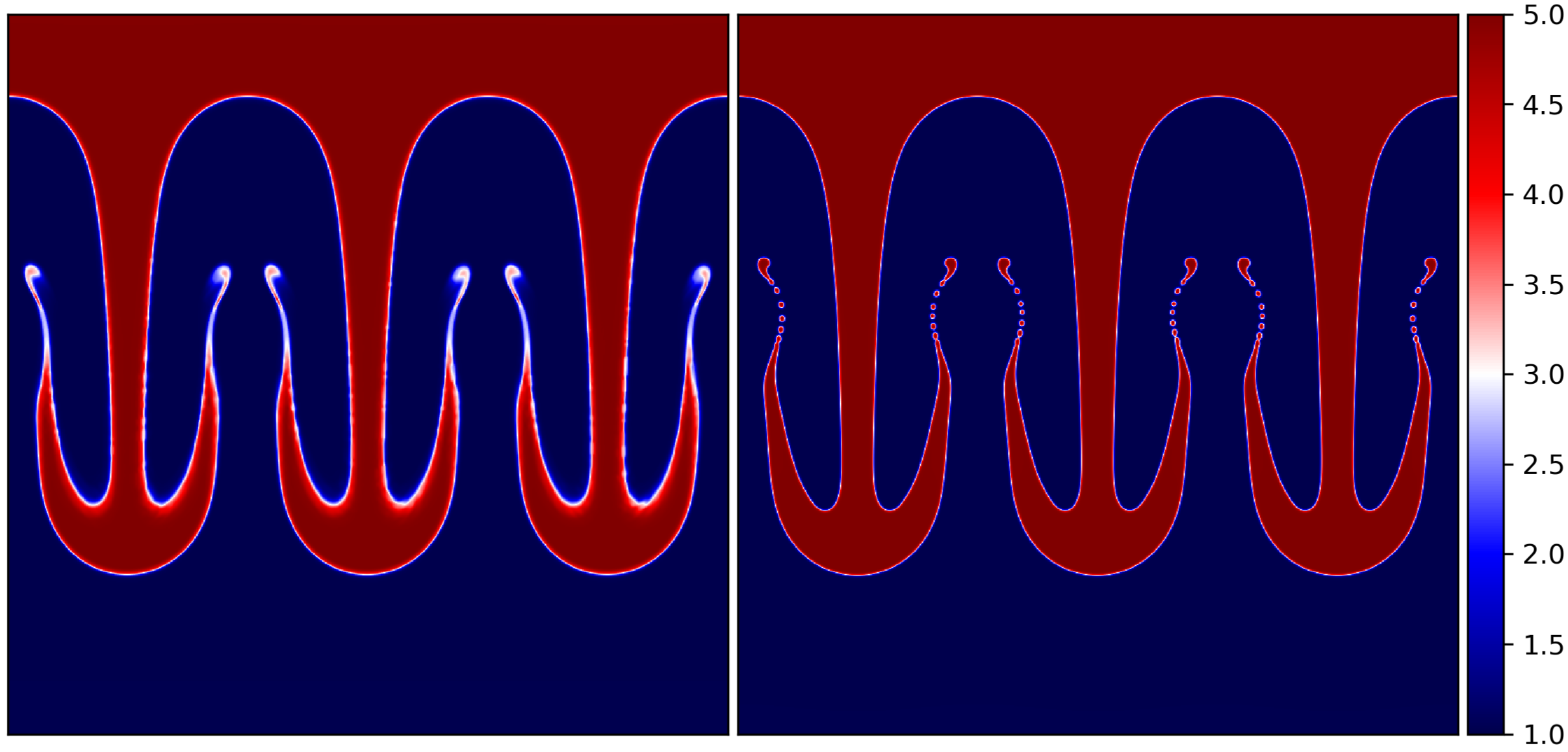}
\caption{Comparison of density field for RT-instability with (right) and without (left) interface regularization terms. \label{fig:RayleighTaylorComparison}}
\end{center}
\end{figure}

A comparison of results achieved using an implicit interface capturing scheme (WENO5Z) and using an explicit phase field method (WENO5Z+CDI) is shown for a Rayleigh-Taylor instability in Figure \ref{fig:RayleighTaylorComparison}. The setup for the case follows the details presented in \cite{collis2024diffuse} and Figure \ref{fig:RayleighTaylorComparison} shows the density field at $t=1.0125$s. As shown, the two approaches have different effects. The implicit interface scheme represents subgrid interfaces with mixing, whereas the explicit interface regularization model will keep interfaces immiscible. However, the interface regularization terms will result in preemptive breakup of under-resolved interfaces even in the absence of surface tension. The applicability of adding interface regularization terms can be dependent on the spatial and temporal scales of the problem of interest.

The interface regularization proposed in this work is an extension of the conservative diffuse interface (CDI) model \cite{Chiu2011,Mirjalili2020,Mirjalili2021,Jain2020} to incorporate multi-component mixtures. The multi-phase multi-component CDI model is designed to capture immiscible interfaces between phases and treat all components within phases as confined scalars \cite{mirjalili2022computational} which can go through intraphase diffusion without restriction. The model form of the multi-phase multi-component CDI model can be written as,

\begin{equation} \label{eq:InterfaceRegularization}
     \mathcal{F_{DI}}(\bm U) = \begin{bmatrix} 
    R^c_{p,x}  \\ 
    u R_{,x} \\ 
    v R_{,x} \\ 
    w R_{,x}  \\ 
    k R_{,x} + \sum_p h_pR_{p,x} 
    \end{bmatrix} \quad 
     \mathcal{G_{DI}}(\bm U) = \begin{bmatrix} 
    R^c_{p,y}  \\ 
    u R_{,y} \\ 
    v R_{,y} \\ 
    w R_{,y}  \\ 
    k R_{,y} + \sum_p h_pR_{p,y} 
    \end{bmatrix} \quad
     \mathcal{H_{DI}}(\bm U) = \begin{bmatrix} 
    R^c_{p,z}  \\ 
    u R_{,z} \\ 
    v R_{,z} \\ 
    w R_{,z}  \\ 
    k R_{,z} + \sum_p h_pR_{p,z} 
    \end{bmatrix}
\end{equation}
with,
\begin{equation} \label{eq:MPMC_CDI}
   R^c_{p,x} = -\frac{Y_p^c}{Y_p} \rho_p  \Gamma\left[\epsilon\frac{\partial \phi_p}{\partial x} - (\phi_p - \phi_{min})(1-(\phi_p-\phi_{min}))\frac{\frac{\partial \phi_p}{\partial x}}{|\vec{\nabla}\phi_p|}\right], 
\end{equation}
and $k = \frac{1}{2}\bm u \cdot \bm u$ as the kinetic energy, $R_{p,x} = \sum_c R_{p,x}^c$, $R_{,x} = \sum_p R_{p,x}$, and $\rho_p$ and $h_p$ are defined using the EOS and governing mixing rules of the given phase $p$. Additionally, $\Gamma$, $\epsilon$, and $\phi_{min}$ are user-set parameters. The parameter $\Gamma$ is a velocity scale which determines the speed the interface equilibrates towards a hyperbolic tangent profile. The parameter $\epsilon$ is a length scale and determines the thickness of the interface. The parameter $\phi_{min}$ determines the volume fraction floor which the regularization term relaxes towards. In this work, $\phi_{min} = 10^{-8}$ for all cases. The multi-phase form of Eq. \ref{eq:MPMC_CDI} has been studied extensively and it has been shown that properly choosing $\epsilon \sim \Delta$ and $\Gamma = \max(|u_{i,j,k}|, |v_{i,j,k}|, |w_{i,j,k}|), \text{ } \forall \text{ } i,j,k$ results in bounded volume fraction for a large range of circumstances, including incompressible flows \cite{Mirjalili2020,Mirjalili2021}, compressible flows \cite{Jain2020}, and recently for geometries using generalized curvilinear grids \cite{collis2024diffuse}. However, when numerically solving Eq. \ref{eq:MPMC_CDI} interpolation errors of $\phi_p$ can occur close to the interface since the regularization sharpens the interface towards a hyperbolic tangent profile of thickness $\epsilon$. Interpolations of $\phi_p$, and its derivatives, are needed to calculate the non-linear sharpening term and the interface normal on a cell face. In order to reduce numerical errors during these interpolations, $\phi_p$ can be transformed to an approximate signed-distance function $\psi$ \cite{shukla2014nonlinear,chiodi2017reformulation,waclawczyk2015consistent,Jain2022ACDI} using,

\begin{equation} \label{eq:psi_transformation}
    \psi = \epsilon \ln{\frac{\phi + \delta}{1-\phi +\delta}}
\end{equation}
where $\delta$ is a very small number (taken as $\delta = 10^{-100}$ in this work). Near the interface the $\psi$ field is approximately linear and will not suffer from numerical error associated with interpolation. An analytical reformulation of Eq. \ref{eq:MPMC_CDI} in terms of $\psi$ has been previously completed \cite{Jain2022ACDI} and results in,
\begin{equation} \label{eq:MPMC_ACDI}
   R^c_{p,x} = -\frac{Y_p^c}{Y_p} \rho_p  \Gamma\left[\epsilon\frac{\partial \phi_p}{\partial x} - \frac{1}{4}\left(1-\tanh^2\Big(\frac{\psi}{2\epsilon}\Big)- 4\phi_{min}(1-\phi_{min})\right)\frac{\frac{\partial \psi}{\partial x}}{|\vec{\nabla}\psi|}\right].
\end{equation}

Both models shown in Eq. \ref{eq:MPMC_CDI} and Eq. \ref{eq:MPMC_ACDI} are equivalent on the continuous level, but the numerical advantages of transforming the nonlinear term from $\phi$ to $\psi$ make the transformed model more computationally attractive. As such, the regularization model in Eq. \ref{eq:MPMC_ACDI} will be used for all simulations in this work.

\subsubsection{Numerical implementation of regularization model}

Analogous to the formulation of the ENO-type scheme to satisfy the IEC, the thermodynamic quantities in the regularization flux can be independently interpolated and subsequently averaged to construct a face flux that adheres to the IEC. Accordingly, the regularization flux in Eq. \ref{eq:InterfaceRegularization} is implemented using a second-order skew-symmetric split-form centered scheme that satisfies the IEC while preserving kinetic energy and entropy (KEEP) \cite{Jain2022KEEP}  

As an example, the numerical flux for the regularization terms on a cell face in the x-direction can be written as,
\begin{align} \label{eq:SkewSymmetricSplitting_Regularization}
  \widehat{\mathcal{F_{DI}}(\bm U)}^{(i\pm 1/2)} = \begin{bmatrix} 
    \widehat{R^c_{p,x}}^{(i\pm 1/2)}  \\ 
    \overline{u}^{(i\pm 1/2)}\widehat{R_{,x}}^{(i\pm 1/2)} \\ 
    \overline{v}^{(i\pm 1/2)}\widehat{R_{,x}}^{(i\pm 1/2)} \\ 
    \overline{w}^{(i\pm 1/2)}\widehat{R_{,x}}^{(i\pm 1/2)}  \\ 
    \overline{k}^{(i\pm 1/2)}\widehat{R_{,x}}^{(i\pm 1/2)} + \sum_p \widehat{h_pR_{p,x}}^{(i\pm 1/2)} 
    \end{bmatrix} \quad
\end{align}
where
\begin{align} \label{eq:SkewSymmetricSplitting_Regularization_Expanded}
   & \widehat{a_{p,x}}^{(i\pm 1/2)} = \Gamma\left[\epsilon\widehat{\frac{\partial \phi_p}{\partial x}}^{(i\pm 1/2)} - \frac{1}{4}\left(1-\tanh^2\Big(\frac{\overline{\psi}^{(i\pm 1/2)}}{2\epsilon}\Big)-4\phi_{min}(1-\phi_{min})\right)\widehat{\frac{\frac{\partial \psi}{\partial x}}{|\vec{\nabla}\psi|}}^{(i\pm 1/2)}\right]\\
   &\widehat{R^c_{p,x}}^{(i\pm 1/2)} = -\overline{\frac{Y_p^c}{Y_p} \rho_p}^{(i\pm 1/2)}\widehat{a_{p,x}}^{(i\pm 1/2)}\\
   &\widehat{R_{,x}}^{(i\pm 1/2)} = \sum_p\sum_c \widehat{R^c_{p,x}}^{(i\pm 1/2)} \\
   &\overline{k}^{(i\pm 1/2)} = \frac{1}{2}(u^{(i\pm1)}u^{(i)}+v^{(i\pm1)}v^{(i)}+w^{(i\pm1)}w^{(i)}) \\
   &\widehat{h_pR_{p,x}}^{(i\pm 1/2)} = -\overline{h_p\rho_p}^{(i\pm 1/2)}  \widehat{a_{p,x}}^{(i\pm 1/2)}.
\end{align}
In the notation above, $\widehat{(\cdot)}^{(i\pm 1/2)}$ denotes a field which consists of a numerical derivative operation onto a face, and $\overline{(\cdot)}^{(i\pm 1/2)}$ represents an interpolation operation onto a face. The critical part of skew-symmetric split-form, which allows the scheme to satisfy IEC, is keeping all velocity and density interpolations to faces independent. Furthermore, it is critical to interpolate the product of enthalpy and density together, as this will retain a consistent pressure field at the cell face. Details on the formulations for each specific term of the CDI model can be found in the following reference \cite{Jain2022KEEP}, and an example on how to extend split form schemes to higher-order spatial accuracy is shown here \cite{Kuya2021}.

\subsection{Positivity preservation} \label{sec:PositivityPreservation}

Section \ref{sec:spatialDiscretization} describes the numerical methods that provide nearly non-oscillatory solutions for flows with shocks and material interfaces. However, the high-order ENO-type schemes still lead to small oscillations which can result in inadmissible thermodynamic states (e.g. negative internal energy or density) and code failure. The positivity of the speed of sound is a critical metric for simulation robustness. To guarantee a positive sound speed the equation of state can be used where the hyperbolic speed of sound is defined by,

\begin{equation} \label{eq:SpeedOfSound}
    a^2 = \frac{C_P}{\rho \beta C_P - \alpha^2T}
\end{equation}
where $\beta = \frac{1}{\rho}\left(\frac{\partial \rho}{\partial P}\right)_{T,Y}$ and $\alpha = -\frac{1}{\rho}\left(\frac{\partial \rho}{\partial T}\right)_{P,Y}$. For the NASG EOS used in this work \cite{peden2023numerical},
\begin{align}
    \beta = \rho \sum_p\sum_cY_p^c\left(\frac{(\gamma^c_p - 1)C{_v}^c_{p}T}{(P+P{_\infty}_{p}^{c})^2}\right) \\
    \alpha = \rho \sum_p\sum_cY_p^c\left(\frac{(\gamma^c_p - 1)C{_v}^c_{p}}{(P+P{_\infty}_{p}^{c})}\right).
\end{align}

For a single-phase flow where $P{_\infty}_{p}^{c} = 0$ $\forall$ $c, p$, the temperature, pressure, and density must be positive to keep the square of the sound speed positive. For a multi-phase flow it is possible for the squared speed of sound to be positive even if mixture pressure is negative. Physical situations like cavitation can result in negative pressures, and the NASG equation of state used in this work permits this. From Eq. \ref{eq:NASG_T1} and Eq. \ref{eq:NASG_T2}, to guarantee an admissible temperature, both the internal energy and the density of the mixture must be positive. Additionally, a positive temperature and pressure will guarantee positive phasic internal energies and densities from Eq. \ref{eq:NASG_EOS}. For both single- and multi-phase flows, to guarantee the positivity of the squared speed of sound with the NASG equation of state the following two requirements must hold,

\begin{equation} \label{eq:PositivityReqs1}
        (e - q) > 0 \quad \text{and} \quad (v - b) > 0.
\end{equation}
With these requirements satisfied the solution will have an admissible speed of sound, temperature, pressure, and mixture density. In order to guarantee boundedness for mass fractions and volume fraction between zero and one, a separate check must take place,
\begin{equation} \label{eq:PositivityReqs2}
        (Y^c_p > 0) \quad \forall \quad c, p.
\end{equation}

To ensure a physically admissible solution at all times, three limiters are added to the numerical procedure. In the Godunov approach described in Section \ref{sec:GodunovApproach}, an interpolation limiter is added after the ENO-type high-order interpolation to the face, and a flux limiter is added after the approximate Riemann solver. Additionally, a flux limiter is added after the interface regularization flux described in Section \ref{sec:DiffuseInterfaceTerms} due to the addition of the sharpening term.  

\subsubsection{Interpolation limiter}

After high-order ENO-type interpolation (step 4 of the Godunov approach), the boundedness of the mass fractions is achieved by limiting the mass fractions using an approach described in \cite{BAUMGART2024113199}.
As an example, the procedure for interpolation of the mass-fraction on the minus side of the $(i+1/2)$ face will be described. First, the mass fraction field left out of the characteristic projection is individually interpolated to the cell face using an ENO-type scheme. Then, the mass fractions can be limited with the following steps,

\begin{enumerate}
    \item If $\sum_c\sum_p Y^c_p{_{(i+1/2)}^-} > 1$
    \begin{enumerate}
        \item $\Sigma^- = \sum_c\sum_p\left(\min\left(\min\left(Y^c_p{_{(i+1/2)}^-}, Y^c_p{_{(i)}}\right) - Y^c_p{_{(i+1/2)}^-}\right)\right)$
        \item $\epsilon^c_p = \left(\frac{1-\sum_c\sum_p Y^c_p{_{(i+1/2)}^-}}{\Sigma^-}\right)\left(\min\left(Y^c_p{_{(i+1/2)}^-}, Y^c_p{_{(i)}} - Y^c_p{_{(i+1/2)}^-}\right)\right)$
    \end{enumerate}
        \item If $\sum_c\sum_p Y^c_p{_{(i+1/2)}^-} < 1$
    \begin{enumerate}
        \item $\Sigma^+ = \sum_c\sum_p\left(\max\left(\max\left(Y^c_p{_{(i+1/2)}^-}, Y^c_p{_{(i)}}\right) - Y^c_p{_{(i+1/2)}^-}\right)\right)$
        \item $\epsilon^c_p = \left(\frac{1-\sum_c\sum_p Y^c_p{_{(i+1/2)}^-}}{\Sigma^+}\right)\left(\max\left(Y^c_p{_{(i+1/2)}^-}, Y^c_p{_{(i)}} - Y^c_p{_{(i+1/2)}^-}\right)\right)$
    \end{enumerate}
    \item $Y^c_p{_{(i+1/2)}^-} = Y^c_p{_{(i+1/2)}^-} + \epsilon^c_p$.
\end{enumerate}

The mass fraction limiter for the positive side of the $(i+1/2)$ cell face is done symmetrically. One of the advantages of this limiter is that for multi-component systems it will ensure that interpolation error does not impact inert species in the stencil and only those with varying profiles. It will also spread out error between multiple species instead of forcing all interpolation errors on the species left out of the interpolation.

Additionally, the positivity of pressure and temperature can be independently checked after the ENO-type interpolation and mass-fraction limiter. If a field is inadmissible, a first-order interpolation (cell-centered value) is taken as the corresponding left/right state to build the flux in the Riemann solver. For example,

\begin{align}
    P_{i+1/2}^- = P_{i} (1-\theta_P) + P_{(i+1/2)}^- \theta_P \\
    T_{i+1/2}^- = T_{i} (1-\theta_T) + T_{(i+1/2)}^- \theta_T.
\end{align}

Note that unlike works with similar limiters \cite{wong2022positivity,bezgin2402jax}, in this work $\theta$ and $\epsilon^c_p$ are kept independent for each field after interpolation, allowing for a less dissipative positivity-preserving interpolation. 

Once all fields in the ENO-type interpolation have been limited, the state on the cell face can be used to build admissible conserved variables (step 5 of Godunov approach) and a Riemann solver can be used to find a flux (step 6 of the Godunov approach).

\subsubsection{Flux limiter}

After using the Riemann solver to find a flux on the cell face, a flux limiter is used to guarantee that the state at the following time step is admissible \cite{wong2022positivity,bezgin2402jax}. An outline for limiting the advection flux in the x-direction with an approximate HLLC Riemann solver is shown below.  
\begin{enumerate}
    \item Calculate $\mathcal{F}^{\bm {HLLC}}(\bm U_{(i+1/2)}^-,\bm U_{(i+1/2)}^+)$
    \item Use pseudo-time integration within the RK sub-step to check that $\mathcal{F}^{\bm {HLLC}}(\bm U_{(i+1/2)}^-,\bm U_{(i+1/2)}^+)$ is admissible. Here $D$ is the spatial dimension of the simulation and $RK_0, RK_1,$ and $RK_2$ are coefficients for a given SSP RK sub-step as shown in Eq. \ref{eq:SSPRK3}. 
    \begin{enumerate}
        \item $\Tilde{\bm U}_{i}^{(RKstep+1)} = RK_0\bm U_{i}^{(n)} + RK_1\bm U_{i}^{(RKstep)} + 2RK_2\Delta tD\mathcal{F}^{\bm {HLLC}}$
        \item $\Tilde{\bm U}_{i+1}^{(RKstep+1)} = RK_0\bm U_{i+1}^{(n)} + RK_1\bm U_{i+1}^{(RKstep)} - 2RK_2\Delta tD\mathcal{F}^{\bm {HLLC}}$
    \end{enumerate}
    \item If Eqs. \ref{eq:PositivityReqs1} and \ref{eq:PositivityReqs2} hold for both $\Tilde{\bm U}_{i}^{(RKstep+1)}$ and $\Tilde{\bm U}_{i+1}^{(RKstep+1)}$, $\mathcal{F}^{\bm {HLLC}}(\bm U_{(i+1/2)}^-,\bm U_{(i+1/2)}^+)$ is admissible. Otherwise, a first-order positivity-preserving flux, $\mathcal{F}^{\bm {HLLC}}(\bm U_{(i)},\bm U_{(i+1)})$, is used.
\end{enumerate}
Step $3$ of the outline above can be augmented with a blending operation between $\mathcal{F}^{\bm {HLLC}}(\bm U_{(i+1/2)}^-,\bm U_{(i+1/2)}^+)$ and $\mathcal{F}^{\bm {HLLC}}(\bm U_{(i)},\bm U_{(i+1)})$. Initial tests using blending did not show a noticeable improvement in the results. In order to reduce the complexity and cost of the positivity preserving routine, blending was not used in this work. It is also useful to observe that checking the solution satisfies  the constraints in Eqs. \ref{eq:PositivityReqs1} and \ref{eq:PositivityReqs2} only requires the conserved vector $\bm U$ without the evaluation of the NASG EOS. For this work, the NASG EOS is analytical for all fields, but for more complex systems, including those with non-constant specific heats, an expensive iteration procedure could be required to evaluate pressure and temperature \cite{cook2009enthalpy}. Evaluating Eqs. \ref{eq:PositivityReqs1} and \ref{eq:PositivityReqs2} remains inexpensive, even when the EOS becomes more complex.

After calculating the diffuse interface flux from Section \ref{sec:DiffuseInterfaceTerms}, the diffuse interface flux can be deemed admissible by using the same flux-limiter algorithm described above. In step $1$, replace $\mathcal{F}^{\bm {HLLC}}$ with $\mathcal{F}^{\bm {HLLC}}+\mathcal{F}^{\bm {DI}}$, where $\mathcal{F}^{\bm {HLLC}}$ is the admissible flux from the advection term. If the diffuse-interface flux is inadmissible it is not added to the solution. Restricting the use of the diffuse-interface flux is infrequent and does not result in any visible smearing/mixing of immiscible multi-phase interfaces for the simulations in Section \ref{sec:NumericalResults}.
\subsection{Temporal integration}

After discretization, the system is expressed as a set of ordinary differential equations which can be integrated in time. In this work we use a third-order strong-stability preserving (SSP) Runge-Kutta method expressed below \cite{gottlieb2001strong},
\begin{equation} \label{eq:SSPRK3}
\begin{split}
    &u^{(1)} = u^n + \Delta t L[u^n] \\
    &u^{(2)} = \frac{3}{4}u^n + \frac{1}{4}u^{(1)} + \frac{1}{4}\Delta t L[u^{(1)}] \\
    &u^{n+1} = \frac{1}{3}u^n + \frac{2}{3}u^{(2)} + \frac{2}{3}\Delta tL[u^{(2)}] 
\end{split}
\end{equation}
where $L[\cdot]$ is the operator to evaluate the numerical approximation of the spatial operations. The superscripts with parenthesis, (1) and (2), denote the first and second denote the sub-steps within time step $n$. The superscripts without parenthesis, $n$ and $n+1$, denote the current time step $n$ as well as the subsequent time step $n+1$. Additionally, the time step size $\Delta t$ is given by either the advection or diffusion CFL criterion expressed as,
\begin{equation}
    \Delta t = \frac{CFL}{\max_{i=1:3}{(\frac{|u_i+a|}{\Delta x_i}, \frac{4\mu/\rho}{\Delta x_i^2}})}
\end{equation}
where $0 < CFL \leq 1$ and $a$ is the hyperbolic speed of sound given by Eq. \ref{eq:SpeedOfSound}.

\subsection{Boundary Conditions: Navier-Stokes characteristic boundary conditions} \label{sec:BoundaryConditions}

The Navier-Stokes characteristic boundary conditions \cite{thompson1990time,poinsot1992boundary}(NSCBC) are used for non-reflecting inflow and outflow conditions. Details on single-phase multi-component uses of the NSCBC conditions can be found in \cite{okong2002consistent}.
Extension of the classic multi-component mixing rules to include multi-phase multi-component flows for the NASG equation of state has been completed \cite{peden2023numerical} and is used in this work.

\subsection{High performance computing}
The implementation of the proposed models and numerical schemes was completed in the highly parallel Hypersonic Task-based Regent (HTR) solver \cite{DiRenzo2020,di2021htr,di2022htr}. HTR is a task-based solver built on the Legion runtime which provides portability to run distributed simulations on both CPUs and GPUs \cite{bauer2012legion}. 

\section{Numerical results} \label{sec:NumericalResults}
The formulation presented in Section \ref{sec:PhysicalModel} is written in a general form and is applicable to flows ranging from single-component single-phase to multi-phase multi-component without changing the numerical scheme, adding redundant equations, or changing the equilibrium assumptions of the four-equation model. The applicability of the formulation for these regimes will be shown with simulations focused on single-phase flows in \ref{sec:SinglePhaseTests}, single-phase multi-component tests in Section \ref{sec:MulticomponentTests}, multi-phase single-component tests in Section \ref{sec:MultiphaseTests}, and multi-phase multi-component  tests in Section \ref{sec:MultiphaseMulticomponentTests}. All results, unless otherwise specified, were created using WENO5Z \cite{borges2008improved} for spatial interpolation and a temporal CFL of $0.5$.   

\subsection{Multi-component tests} \label{sec:MulticomponentTests}
In this section two single-phase multi-component simulations are shown without interface regularization to verify the applicability of the framework for multi-component gaseous flows. The first test is a shock-bubble interaction between helium and and air bubble and is compared against an experiment \cite{haas1987interaction}. The second test is a single-mode Richtmyer-Meshkov instability and compared to past computational studies \cite{terashima2009front,hoppe2022alpaca}.

\subsubsection{Shock-bubble interaction: air-helium bubble}

\begin{figure}[hbt!]
\begin{center}
\includegraphics[width=0.6\textwidth]{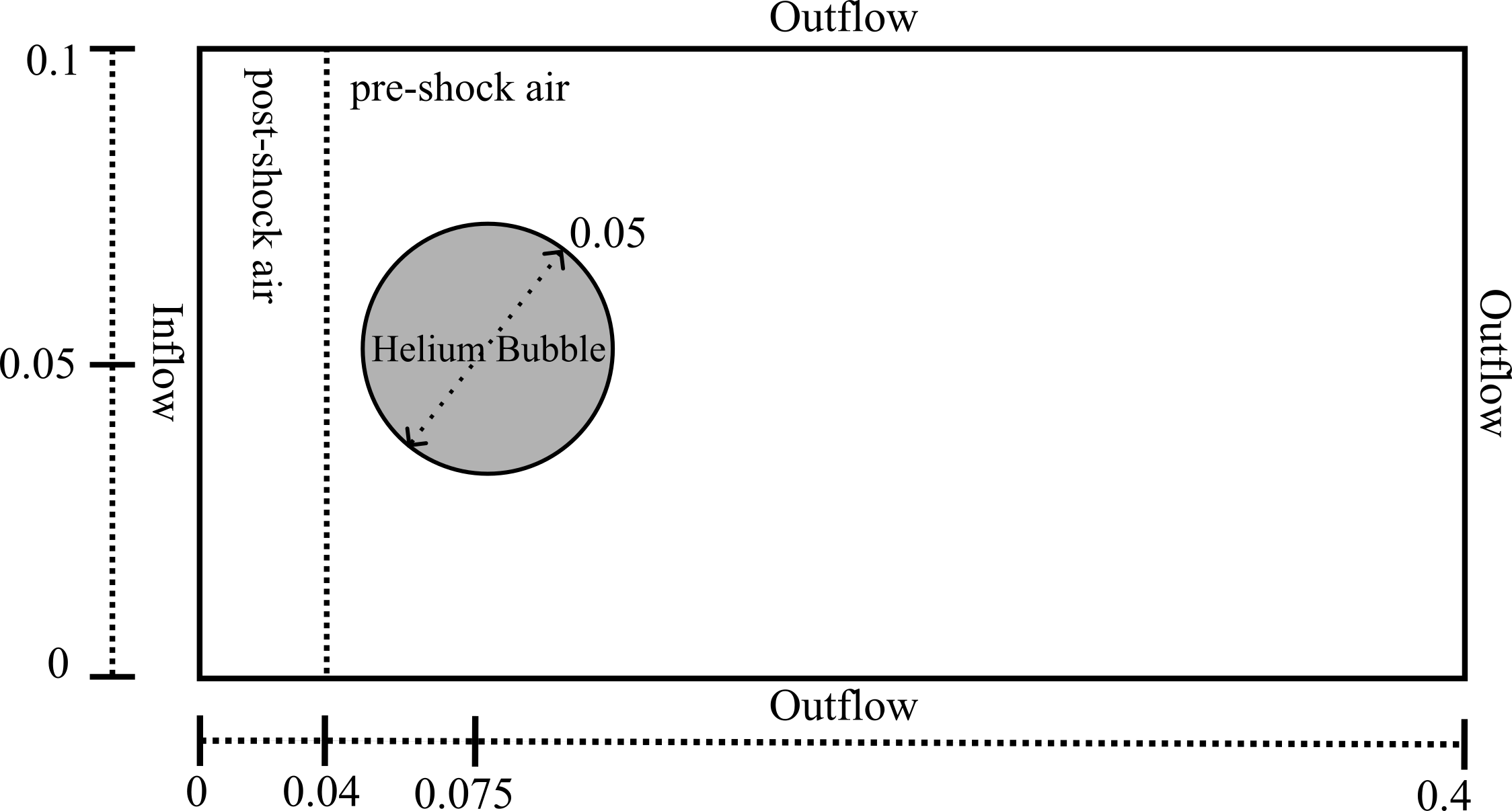}
\caption{Shock-bubble interaction between air and helium. All units are in meters.\label{fig:ShockBubble-AirHeliumSchematic}}
\end{center}
\end{figure}

The interaction of a shock in air with a helium bubble is a well-documented test case for multi-component flows. The schematic showing the setup for this case is shown in Figure  \ref{fig:ShockBubble-AirHeliumSchematic}. The initial thermodynamic state as well as the component viscosity and Schmidt numbers used in this test are shown in Table \ref{tab:ShockBubble-AirHeliumSetup}. Lastly, a mixture Prandtl number of 0.71 is used to define the heat conduction and a spatial resolution of ($4096 \times 1024$) was used.

\begin{table}[hbt!]
    \centering
    \begin{tabular}{l c c c c c }
    \hline
     Material &  $\rho$ [kg/$\text{m}^3$] & u [m/s] & P [Pa] & $\mu$ [Pa-s] & $Sc$ [-]
    \\[1mm]
    \hline
    Helium  & 0.166 & 0.0 & 101325.0 & $1.96\times 10^{-5}$ & $0.70$
    \\[1mm]
    pre-shock air  & 1.18 & 0.0 & 101325.0 & $1.81\times 10^{-5}$ & $1.0$
    \\[1mm]
    post-shock air  & 1.624 & 115.65 & 159050.0 & $1.81\times 10^{-5}$ & $1.0$
    \\[1mm]
    \hline
    \\
    \end{tabular}
    \caption{Initial conditions for 2D shock-bubble interaction between air and helium}
    \label{tab:ShockBubble-AirHeliumSetup}
\end{table}

\begin{figure}[hbt!]
\begin{center}
\includegraphics[width=\textwidth]{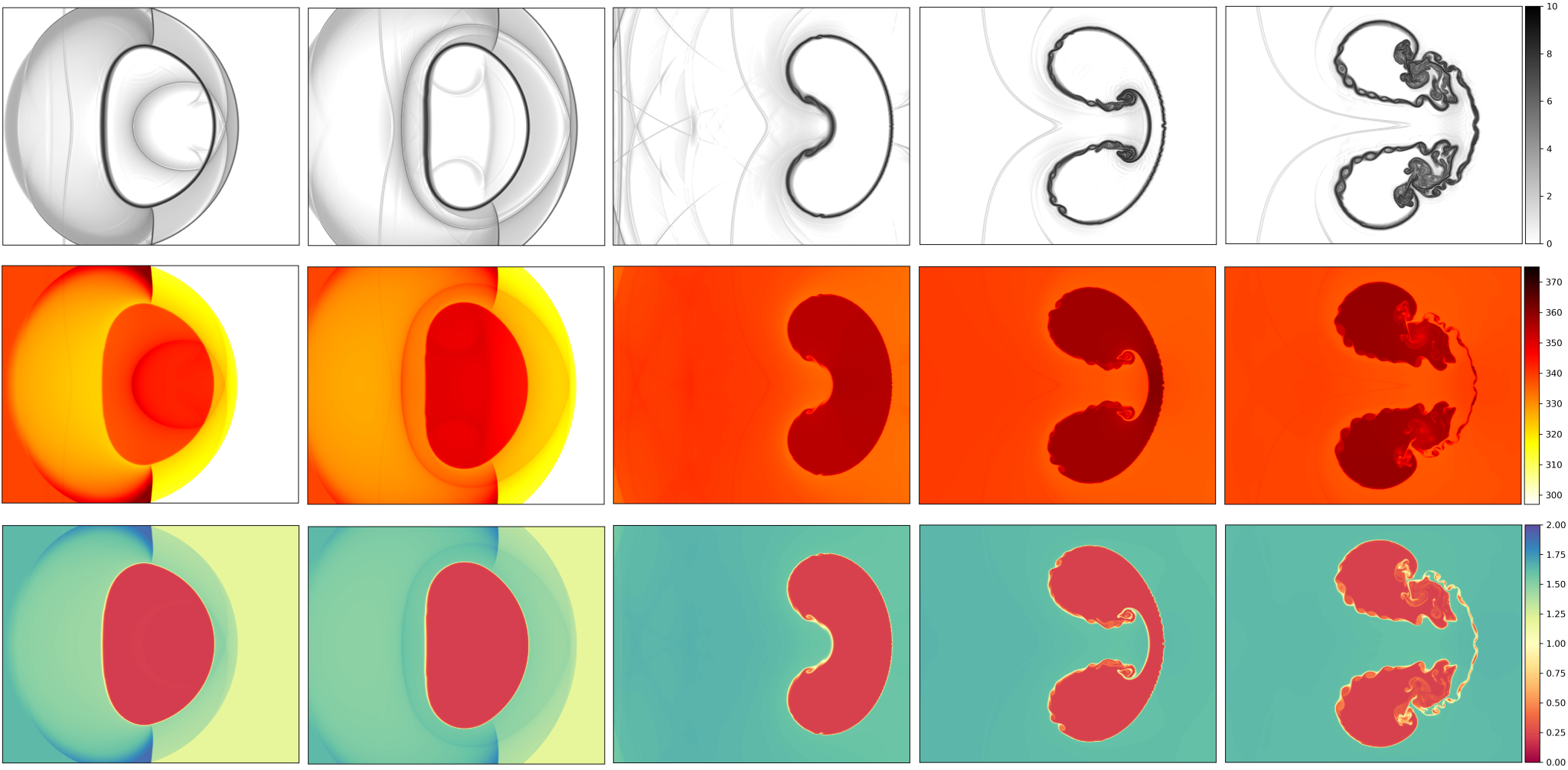}
\caption{Shock-bubble interaction between air and helium. First row:  experiment \cite{haas1987interaction} (not included in arXiv due to limited copyright). Second row: numerical Schlieren $\ln\left(\frac{||\nabla \rho||}{\rho}\right)$. Third row: temperature. Final row: density. From left to right the snapshots are taken at times of $72, 102, 245, 427$, and $674 \mu$s after the shock impacts the helium cylinder. \label{fig:ShockBubble-AirHeliumComparison}}
\end{center}
\end{figure}

For this case the shock-wave speed is 423 m/s and for the setup shown in Figure  \ref{fig:ShockBubble-AirHeliumSchematic}, the shock impacts the helium bubble after 23.6 $\mu$s. As this case is between two miscible species the interface regularization term is not active and species mass diffusion is present between the air and helium. Figure  \ref{fig:ShockBubble-AirHeliumComparison} compares the evolution of the helium bubble with an experiment \cite{haas1987interaction} at times after the shock first makes contact with the helium bubble. All temporal snapshots show good visual agreement between the numerical simulation and the experiment. 

\subsubsection{Single-mode Richtmyer-Meshkov instability}

\begin{figure}[hbt!]
\begin{center}
\includegraphics[width=0.75\textwidth]{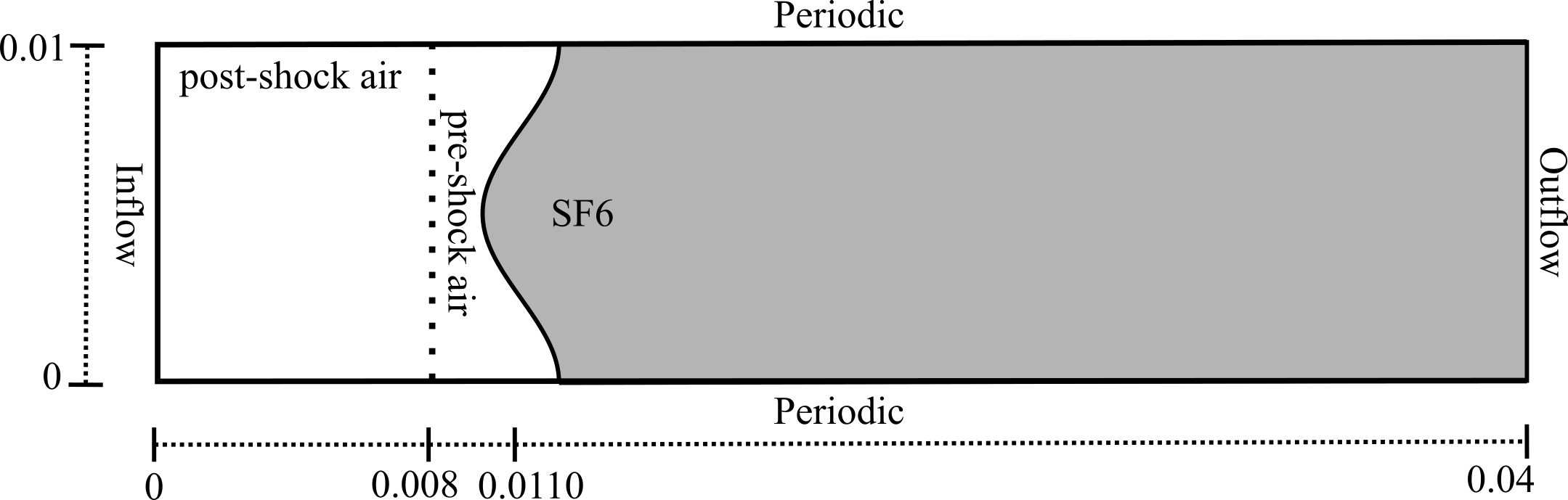}
\caption{Schematic of initial condition for single-mode Rightmyer-Meshkov instability between air and SF6.\label{fig:SingleModeRMISchematic}}
\end{center}
\end{figure}

We consider a 2D single-mode Richtmyer-Meshkov simulation following the computational setup used in past works for additional code-to-code verification \cite{terashima2009front, hoppe2022alpaca}. The thermodynamic quantities used in this test are shown in Table \ref{tab:RMI-SingleModeSetup}. The test consists of a Mach 1.24 shock in air passing through a perturbed interface of SF6 and details on the initial condition are shown in Figure  \ref{fig:SingleModeRMISchematic}.

\begin{table}[hbt!]
    \centering
    \begin{tabular}{l c c c c c }
    \hline
     Material &  $\rho$ [kg/$\text{m}^3$] & u [m/s] & P [Pa] & $\mu$ [Pa-s] & $Sc$ [-]
    \\[1mm]
    \hline
    SF6  & 6.06 & 0.0 & 101325.0 & $1.5\times 10^{-5}$ & $1.66$
    \\[1mm]
    pre-shock air  & 1.18 & 0.0 & 101325.0 & $1.81\times 10^{-5}$ & $1.0$
    \\[1mm]
    post-shock air  & 1.66 & 125.3 & 164957.0 & $1.81\times 10^{-5}$ & $1.0$
    \\[1mm]
    \hline
    \\
    \end{tabular}
    \caption{Initial conditions for 2D single-mode Richtmyer-Meshkov instability}
    \label{tab:RMI-SingleModeSetup}
\end{table}

\begin{figure}[hbt!]
\begin{center}
\includegraphics[width=1\textwidth]{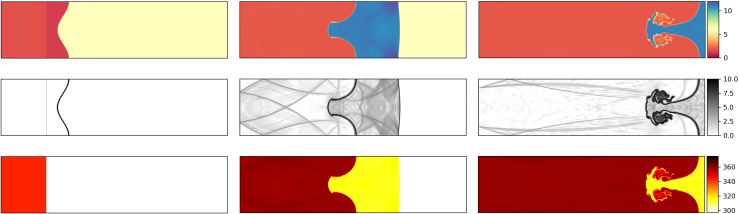}
\caption{Single-mode Rightmyer-Meshkov instability between air and SF6 over time. First row: density. Second row: numerical Schlieren $\left(\ln\left(\frac{||\nabla \rho||}{\rho}\right)\right)$. Last row: temperature. The snapshots are taken at rescaled times $t^*$ from left to right of $0, 3.2$, and $10$. \label{fig:SingleModeRMIDensity}}
\end{center}
\end{figure}

\begin{figure}[hbt!]
\begin{center}
\includegraphics[width=0.4\textwidth]{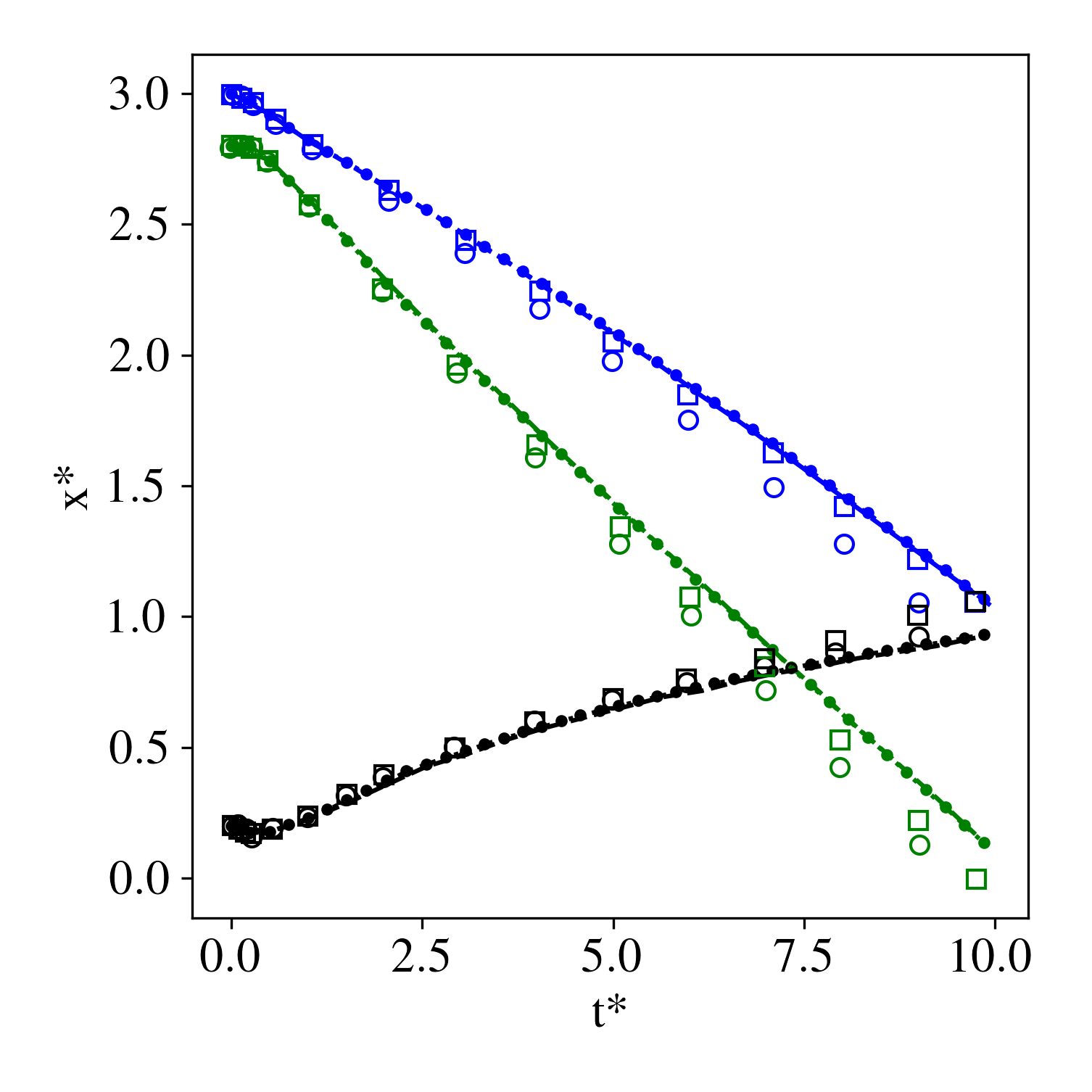}
\caption{Characteristic locations over time for single-mode Rightmyer-Meshkov instability between air and SF6. The labels are listed as location of the (spike, bubble, mixing layer) for the fine resolution ($2048 \times 512$)(\protect\markersixBlue{},\protect\markersixGreen{},\protect\markersix{}), medium resolution ($1024 \times 256$) \textbf{}(\protect\dottedBlue{}{}{},\protect\dottedGreen{}{},\protect\dotted{}), coarse resolution ($512 \times 128$) (\protect\dashedBlue{}{}{},\protect\dashedGreen{}{},\protect\dashed{}), Terishima data \cite{terashima2009front} (\protect\markeroneBlue{},\protect\markeroneGreen{}{},\protect\markerone{}), and Adams data \cite{hoppe2022alpaca}(\protect\markerfourBlue{},\protect\markerfourGreen{}{},\protect\markerfour{}). \label{fig:SingleModeRMICharacteristics}}
\end{center}
\end{figure}

Figure  \ref{fig:SingleModeRMIDensity} shows the development of the RMI over time. After the shock passes through the perturbed interface the characteristic mushroom shape appears and a mixing layer forms between the SF6 and air. The locations of the spike (farthest left point of the interface), the bubble (farthest right point of the interface), and the mixing zone (difference between the spike and bubble location) are reported for three resolutions. These resolutions are ($512 \times 128$), ($1024 \times 256$), and ($2048 \times 512$). Furthermore, the RMI results are compared with previous numerical experiments in Figure \ref{fig:SingleModeRMICharacteristics}. The axis in Figure \ref{fig:SingleModeRMICharacteristics} is scaled to consistently compare with the non-dimensionalization used in previous studies. 

Figure  \ref{fig:SingleModeRMICharacteristics} shows that the three numerical resolutions simulated in this work have converged relative to the locations of the spike, bubble, and mixing zone over time. Additionally, it shows that the results from this work match well with results from past published works \cite{terashima2009front,hoppe2022alpaca}, verifying the implementation and application of the proposed numerical scheme. 

\subsection{Multi-phase tests} \label{sec:MultiphaseTests}
To illustrate the applicability of the regularization model combined with ENO-type schemes to single-component multi-phase flows four tests will be shown in this section. The first is a classic one-dimensional gas-liquid Riemann problem to verify the implementation of the model for multi-phase flows with shocks. The second is an isothermal and inviscid water droplet advection in air to verify the interface equilibrium condition is satisfied. The third case is a shock in air interacting with a water droplet to verify the implementation of the positivity preserving algorithm as this cases will fail without it. Lastly, an inviscid Mach 100 water jet will be simulated to illustrate the robustness of the framework even when applied to unrealistically difficult flows.

\subsubsection{Gas-liquid Riemann problem}

We consider a common gas-liquid Riemann problem which was first analyzed by \cite{liu2003ghost}. This problem is formulated as a model problem for an underwater explosion in which the left state is highly compressed air and the right state is water at atmospheric pressure. The same setup used by \cite{coralic2014finite} is repeated here with a spatial resolution of ($200 \times 1)$ and final time of $t = 0.2$. The initial condition and material parameters are listed in Table \ref{tab:GasLiqSetup}.

\begin{table}[hbt!]
    \centering
    \begin{tabular}{l c c c c c }
    \hline
     Location & Material &  $\rho$ [kg/$\text{m}^3$] & u [m/s] & P [Pa] & $\mu$ [Pa-s]
    \\[1mm]
    \hline
    $-1\leq x < 0.0$ & post-shock air  & 1.241 & 1.0 & 2.753 & $0.0$ 
    \\[1mm]
    $0\leq x < 1.0$ & water & 0.991 & 0.0 & $3.059\times 10{-4}$ & $0.0$ 
    \\[1mm]
    \hline
    \\
    \end{tabular}
    \caption{Initial conditions for 1D gas-liquid Riemann problem}
    \label{tab:GasLiqSetup}
\end{table}

\begin{figure}[hbt!]
\begin{center}
\includegraphics[width=0.65\textwidth]{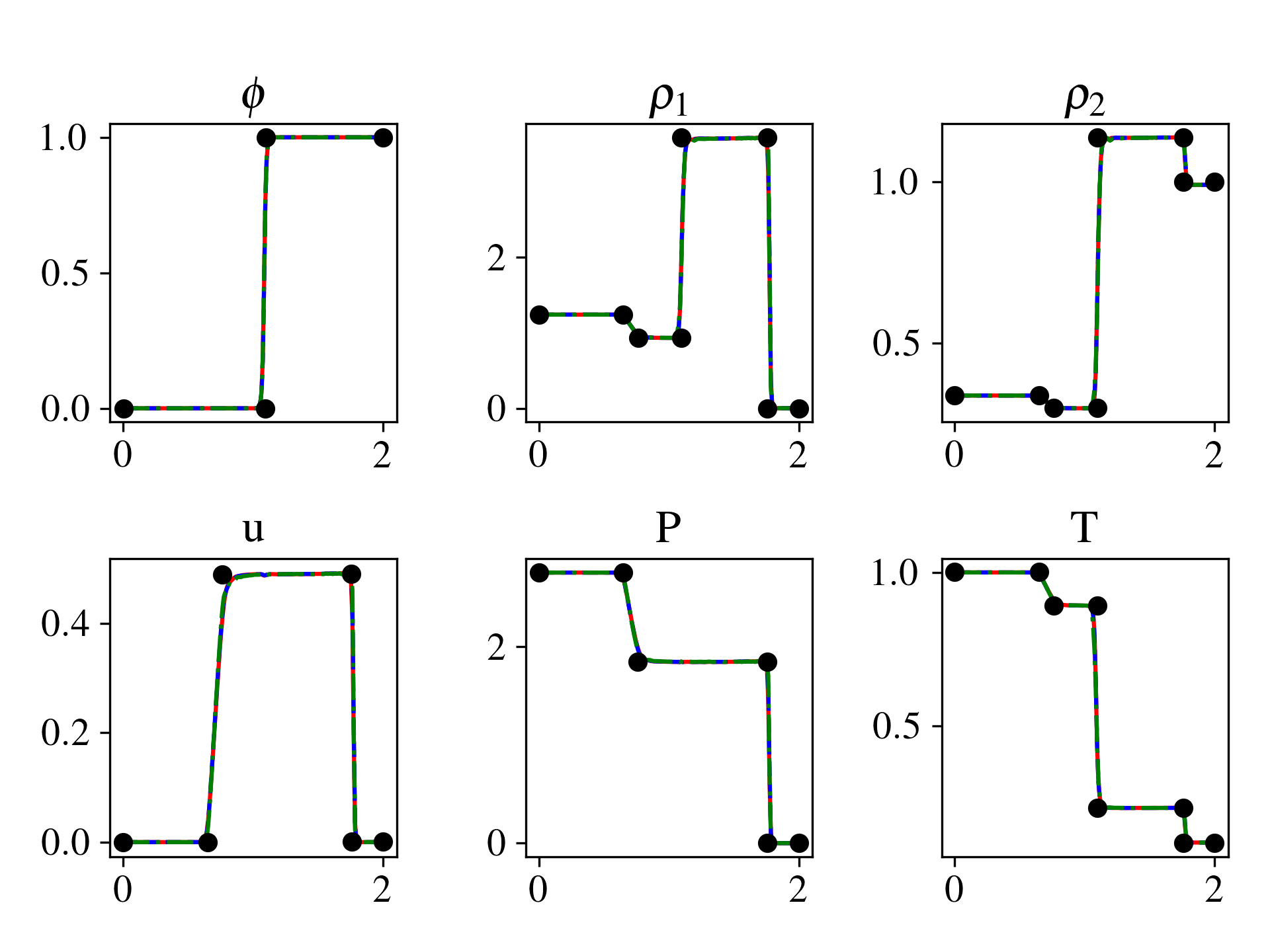}
\caption{Gas-Liquid Riemann problem for three ENO-type schemes. Exact solution: (\protect\markeroneFull{}), WENO5JS \cite{jiang1996efficient} (\protect\fullRed{}), WENO5Z \cite{borges2008improved}(\protect\dashedBlue{}), TENO6 \cite{fu2016family}(\protect\chain{}). \label{fig:GasLiqRiemannProblem}}
\end{center}
\end{figure}

Figure  \ref{fig:GasLiqRiemannProblem} shows the results for TENO6 \cite{fu2016family}, WENO5JS \cite{jiang1996efficient}, and WENO5Z \cite{borges2008improved} compared to the exact solution. All schemes perform well without excessive oscillations in density, pressure, and temperature. Small oscillations can be seen in the phasic density plots which are larger for TENO6 than the WENO-type schemes. Additionally, the regularization of volume fraction keeps the interface at nearly constant thickness for all ENO-type schemes. 

\subsubsection{Inviscid droplet advection} \label{sec:DropAdvection}

As discussed in Section \ref{sec:IEC} it is critical to ensure that the numerical scheme satisfies the interface equilibrium condition (IEC) and does not introduce unphysical oscillations around material interfaces. The standard test for ensuring the scheme satisfies the IEC is an inviscid advection case of a water droplet in air. This test is used extensively for validating the 5-equation model but is more difficult in the 4-equation context. As discussed in Section \ref{sec:IEC}, the 4-equation model requires an oscillation-free temperature field across an isothermal material interface. The additional coupling requires oscillation-free pressure, temperature, and velocity fields. Table \ref{tab:DropAdvectionSetup} outlines the initial setup for this 1D droplet advection case. Two WENO-type schemes and two TENO-type schemes, each coupled to the interface regularization terms described in Section \ref{sec:DiffuseInterfaceTerms}, will be used to verify that the proposed scheme satisfies the IEC for a spatial resolution of ($100 \times 1$). 

\begin{table}[hbt!]
    \centering
    \begin{tabular}{l c c c c c c}
    \hline
    \multicolumn{1}{c}{Interface} & Material &  $\rho$ [kg/$\text{m}^3$] & u [m/s] & P [Pa] & T [K] & $\mu$ [Pa-s]\\
    \hline
    $\phi_l$ = $\frac{1}{2}\left[1+\tanh\left(\frac{0.25 - (x-0.5)}{2\epsilon\Delta}\right)\right]$&  Water & 997 & 5.0 & 101325.0 & 297 & $0.0$ 
    \\[1mm]
    $\phi_g = 1-\phi_l$ & Air & 1.18 & 5.0 & 101325.0 & 297 & $0.0$ 
    \\[1mm]
    \hline
    \\
    \end{tabular}
    \caption{Initial conditions for 2D droplet advection problem}
    \label{tab:DropAdvectionSetup}
\end{table}


\begin{table}[hbt!]
    \centering
    \begin{tabular}{l l l l}
    \hline
    \multicolumn{1}{c}{4-eq IEC ENO-type Scheme} & $\max \left(\frac{|P-P_{exact}|}{P_{exact}}\right)$ &  $\max \left(\frac{|T-T_{exact}|}{T_{exact}}\right)$ & $\max\left( \frac{|u-u_{exact}|}{u_{exact}}\right)$ \\[1mm]
    \hline
    WENO5JS \cite{jiang1996efficient} &  $1.32\times 10^{-11}$ & $6.03\times 10^{-12}$ & $2.71\times 10^{-12}$
    \\[1mm]
    WENO5Z \cite{borges2008improved} &  $1.97\times 10^{-11}$ & $1.04\times 10^{-11}$ & $4.50\times 10^{-12}$
    \\[1mm]
    TENO6 \cite{fu2016family} &  $8.45\times 10^{-10}$ & $3.99\times 10^{-8}$ & $1.12\times 10^{-8}$
    \\[1mm]
    TENO5 \cite{fu2016family} &  $1.60\times 10^{-11}$ & $8.41\times 10^{-12}$ & $6.82\times 10^{-12}$
    \\[1mm]
    \hline
    \end{tabular}
    \caption{Maximum normalized IEC error after one flow-through time for multiple ENO-type schemes using $\bm W = [T,Y^c_p,u,v,w,P]$}
    \label{tab:DropAdvectionIECResults}
\end{table}

\begin{table}[hbt!]
    \centering
    \begin{tabular}{l l l l}
    \hline
    \multicolumn{1}{c}{5-eq IEC ENO-type Scheme} & $\max \left(\frac{|P-P_{exact}|}{P_{exact}}\right)$ &  $\max \left(\frac{|T-T_{exact}|}{T_{exact}}\right)$ & $\max\left( \frac{|u-u_{exact}|}{u_{exact}}\right)$ \\[1mm]
    \hline
    WENO5JS \cite{jiang1996efficient} &  $2.38\times 10^{-3}$ & $1.56\times 10^{-2}$ & $6.03\times 10^{-4}$
    \\[1mm]
    WENO5Z \cite{borges2008improved} &  $1.609\times 10^{-5}$ & $1.37\times 10^{-4}$ & $1.06\times 10^{-5}$
    \\[1mm]
    TENO6 \cite{fu2016family} &  $4.89\times 10^{-4}$ & $3.58\times 10^{-3}$ & $3.19\times 10^{-3}$
    \\[1mm]
    TENO5 \cite{fu2016family} &  $2.47\times 10^{-2}$ & $4.01\times 10^{-1}$ & $9.42\times 10^{-2}$
    \\[1mm]
    \hline
    \end{tabular}
    \caption{Maximum normalized IEC error after one flow-through time for multiple ENO-type schemes using $\bm W = [\rho Y^c_p,u,v,w,P]$ which satisfies IEC for the 5-equation model, but not for the 4-equation. For ease of implementation, the results in this table are run using the Godunov approach without the characteristic decomposition shown in \cite{coralic2014finite}.}
    \label{tab:DropAdvectionNoIECResults}
\end{table}


After advection for one flow-through time, the pressure, temperature and velocity errors are shown in Table \ref{tab:DropAdvectionIECResults} for the proposed IEC ENO-type schemes. The errors for all fields remain near machine precision for all WENO-type schemes and the TENO5 scheme after one period of advection, which verifies that the proposed scheme satisfies the interface equilibrium condition. The TENO6 scheme remains near machine precision for pressure error across the interface but shows a slightly higher error for the velocity and temperature fields due to the sensitivity of TENO6 to the pile-up of round-off errors \cite{fleischmann2019numerical}. For comparison, Table \ref{tab:DropAdvectionNoIECResults} shows the results of the advection test using the basis suggested for the 5-equation model, $\bm W = [\rho Y^c_p,u,v,w,P]$. 
Though using the basis, $\bm W = [\rho Y^c_p,u,v,w,P]$, satisfies IEC for the 5-equation model, enforcing thermodynamic equilibrium to obtain the 4-equation model results in oscillations across the interface many orders of magnitude larger than with the proposed basis of $\bm W = [T,Y^c_p,u,v,w,P]$ for all ENO-type schemes. 




\subsubsection{Shock-droplet interaction: air-water}
\begin{figure}[hbt!]
\begin{center}
\includegraphics[width=0.5\textwidth]{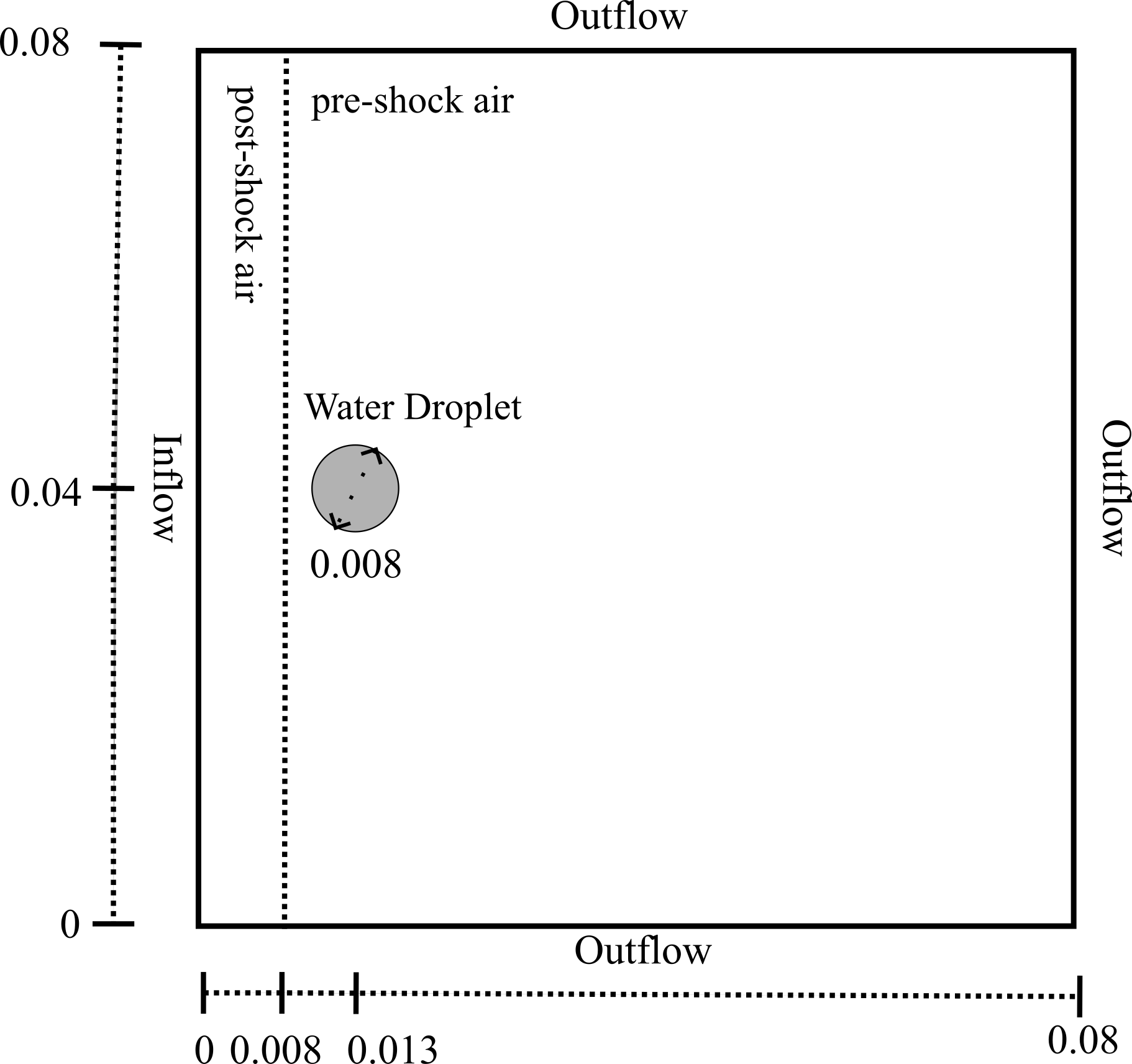}
\caption{Schematic of initial condition for shock-droplet interaction between air and water.\label{fig:ShockDroplet-AirWaterSchematic}}
\end{center}
\end{figure}

For this test we consider a Mach 1.47 shock-wave in air interacting with a water droplet at atmospheric conditions. To remain stable for late times, the test case requires the positivity-preserving algorithm described in Section \ref{sec:PositivityPreservation}. The initial thermodynamic conditions used in this simulation are in Table \ref{tab:ShockDroplet-AirWaterSetup}. A diagram showing the setup of this case is shown in Figure  \ref{fig:ShockDroplet-AirWaterSchematic}. Three spatial resolutions of ($512 \times 512$), ($1024 \times 1024)$, and ($2048 \times 2048$) where used.

\begin{figure}[hbt!]
\begin{center}
\includegraphics[width=0.85\textwidth]{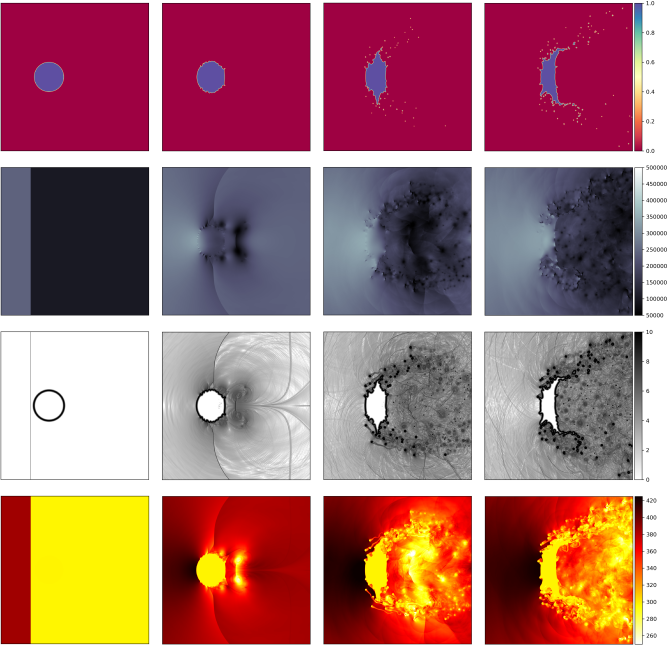}
\caption{Time evolution for shock-droplet interaction between air and water. First row: volume fraction. Second row: pressure. Third row: numerical Schlieren $\left(\ln\left(\frac{||\nabla \rho||}{\rho}\right)\right)$. Fourth row: temperature. Normalized time from left to right: 0, 0.15, 0.4, 0.8. \label{fig:ShockDroplet-AirWaterTimeEvolution}}
\end{center}
\end{figure}

\begin{table}[hbt!]
    \centering
    \begin{tabular}{l c c c c }
    \hline
     Material &  $\rho$ [kg/$\text{m}^3$] & u [m/s] & P [Pa] & $\mu$ [Pa-s]
    \\[1mm]
    \hline
    Water  & 997.0 & 0.0 & 101325.0 & $8.9 \times 10^{-4}$ 
    \\[1mm]
    pre-shock air  & 1.18 & 0.0 & 101325.0 & $1.81\times 10^{-5}$ 
    \\[1mm]
    post-shock air  & 2.14 & 228.007 & 238500.0 & $1.81\times 10^{-5}$ 
    \\[1mm]
    \hline
    \\
    \end{tabular}
    \caption{Initial conditions for 2D shock-bubble: water-air problem}
    \label{tab:ShockDroplet-AirWaterSetup}
\end{table}

\begin{figure}[hbt!]
\begin{center}
\includegraphics[width=0.5\textwidth]{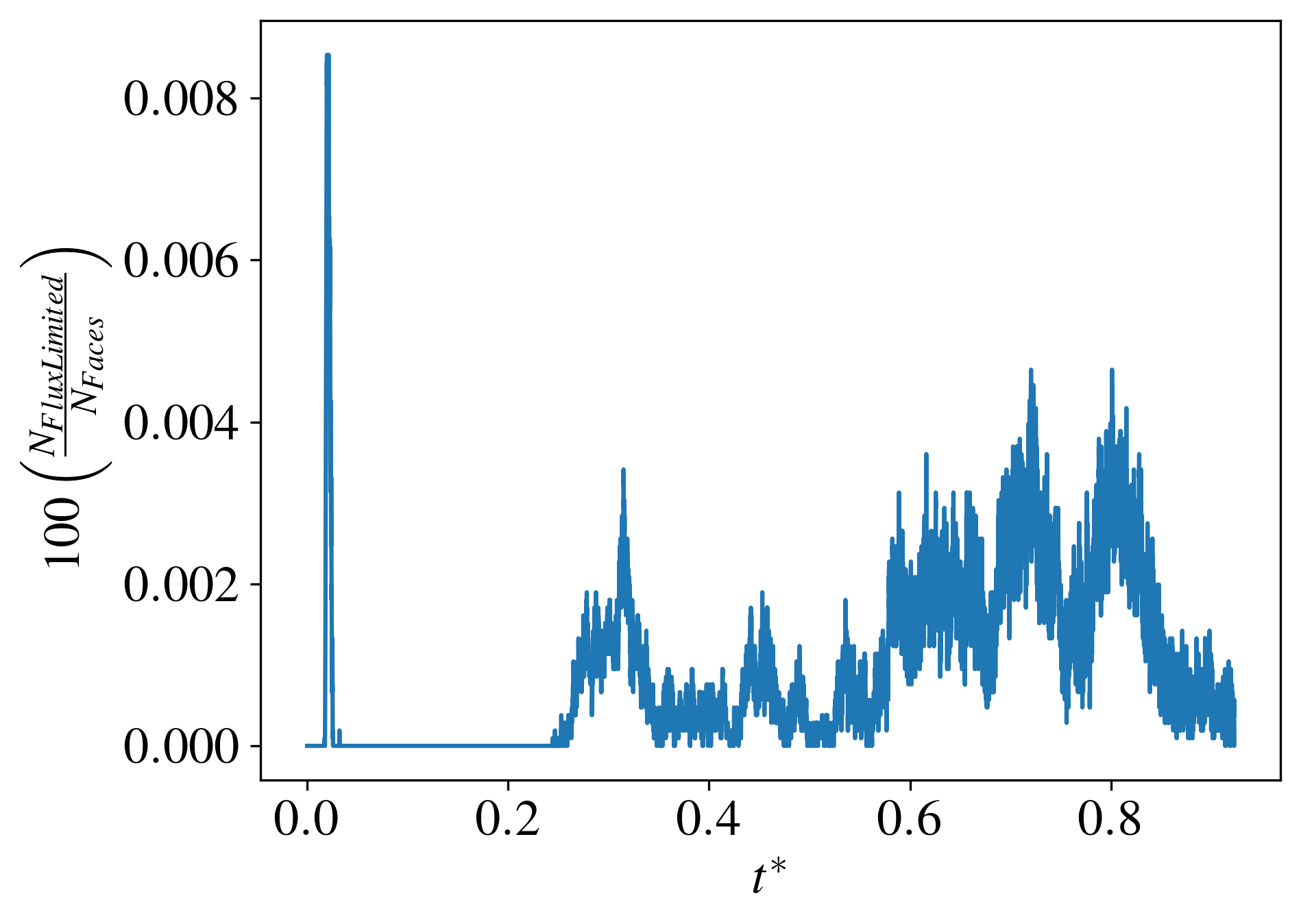}
\caption{Percentage of faces which required flux limiter over time for medium resolution ($1024 \times 1024$) of the shock-droplet interaction between water and air. \label{fig:ShockDroplet-LimiterEvolution}}
\end{center}
\end{figure}

Figure  \ref{fig:ShockDroplet-AirWaterTimeEvolution} shows the volume fraction, pressure, temperature, and Schlieren throughout time. 
After the shock travels through the water droplet a rarefaction wave is formed at the trailing edge. The corresponding pressure wave will travel towards the front of the droplet and create a low pressure zone at the trailing edge. In order to simulate through this point in the simulation the positivity limiter was active for both the WENO5Z interpolation as well as the flux limiter. Figure  \ref{fig:ShockDroplet-LimiterEvolution} shows the activation of the flux-limiter over time. In general, the activation of the limiter is very sparse in the domain, with a maximum percentage of activation for $0.008\%$ of faces. The early sharp rise occurs when the initial rarefaction wave leaves the trailing edge of the drop. These pressure waves will reflect from within the droplet with decreasing magnitude throughout the simulation. Later, a wake region forms behind the water droplet and shear forces deform the droplet creating thin features, resulting in spatially dispersed activation of the limiter. To keep the interface finite and resolvable during deformation of the droplet, the CDI model turns the sheared thin features into small secondary droplets. The ability of the CDI model to form droplets can be interpreted as numerical surface tension. As this problem does not include physical surface tension (Weber number is infinity) the interface deformation and formation of interfacial features smaller than the grid-size will exist at any resolution. As shown in Figure  \ref{fig:ShockDroplet-AirWaterTimeEvolution} the CDI model will represent these unresolved features as immiscible droplets of finite size. Since the CDI model is locally conservative, these droplets do not disappear with time (as opposed to other popular interface capturing schemes which are only globally conservative) and instead are transported throughout the simulation. Without the CDI model the unresolved breakup of the main drop would be modeled by the implicit diffusion of the numerical scheme and result in mixing of immiscible phases.

\begin{figure}[hbt!]
\begin{center}
\includegraphics[width=0.5\textwidth]{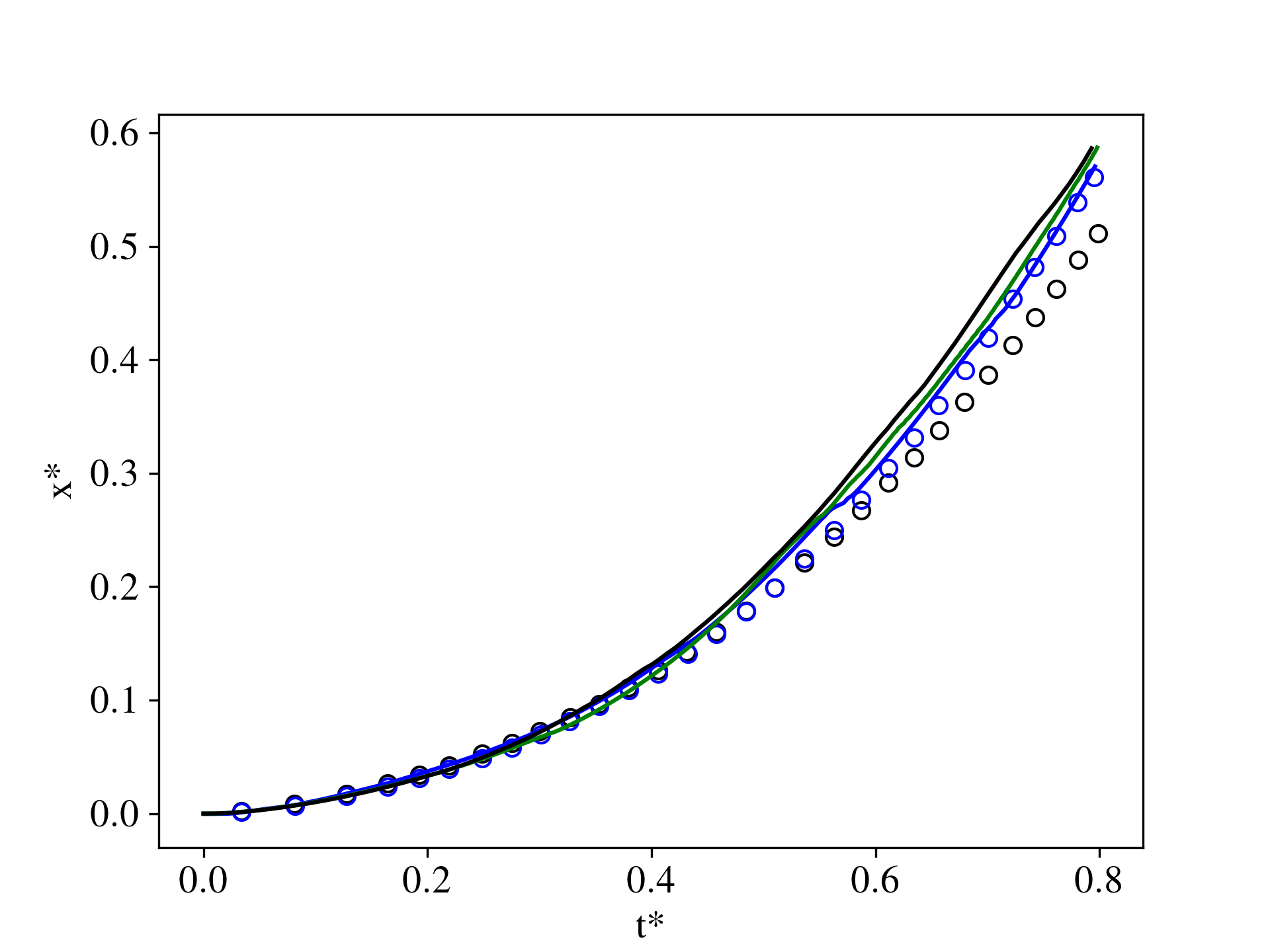}
\caption{Evolution of centroid over time for shock-droplet case. Coarse resolution ($512 \times 512$) (\protect\fullBlue{}), medium resolution ($1024 \times 1024$) (\protect\fullGreen{}), fine resolution ($2048 \times 2048$) (\protect\full{}), Colonius data \cite{meng2015numerical} (\protect\markerone{}), and Man Long data \cite{wong2022positivity}(\protect\markeroneBlue{}).\label{fig:ShockDroplet-CentroidEvolution}}
\end{center}
\end{figure}

The center of mass of the main water droplet is tracked throughout time and compared to three resolutions simulated in this work, as well as with two past computational simulations. These results are reported in Figure  \ref{fig:ShockDroplet-CentroidEvolution} and show good agreement between all cases.

\subsubsection{Mach 100 water column in air}

The final single-component multi-phase simulation we consider is the injection of a Mach 100 water column into ambient air. The test shows the applicability of the framework to cover very large density ratios and strong shocks. The initial conditions for this test are given in Table \ref{tab:Mach100WaterColumnSetup} and a spatial resolution of $(2048\times1025)$ was used with a constant time step of $7.5\times 10^{-4}\mu$s.
\begin{table}[hbt!]
    \centering
    \begin{tabular}{l c c c c c }
    \hline
     Location & Material &  $\rho$ [kg/$\text{m}^3$] & u [m/s] & P [Pa] & $\mu$ [Pa-s]
    \\[1mm]
    \hline
    $x=0$ and $|y| < 0.5$ & 
    Water  & 997.0 & 150000.0 & 101325.0 & $0.0$ \\[1mm]
    otherwise  & Air  & 1.18 & 0.0 & 101325.0 & $0.0$ 
    \\[1mm]
    \hline
    \\
    \end{tabular}
    \caption{Initial condition for 2D Mach 100 water column}
    \label{tab:Mach100WaterColumnSetup}
\end{table}

\begin{figure}[hbt!]
\begin{center}
\includegraphics[width=0.85\textwidth]{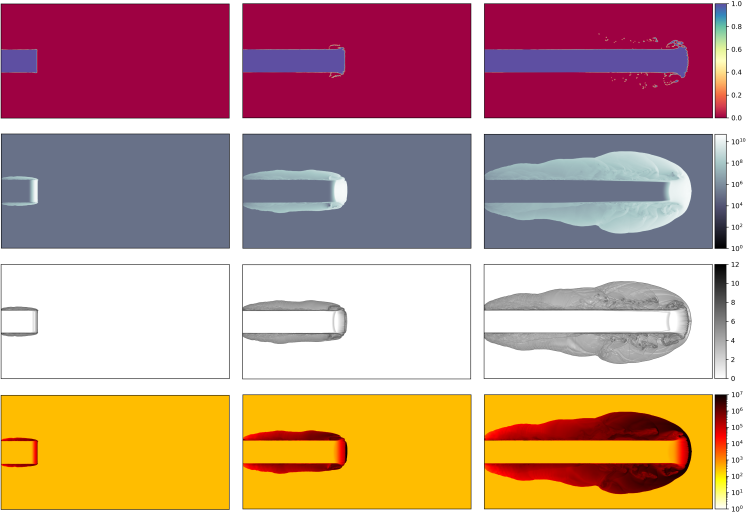}
\caption{Time evolution for Mach 100 water jet into air. First row: volume fraction. Second row: pressure. Third row: numerical Schlieren $\left(\ln\left(\frac{||\nabla \rho||}{\rho}\right)\right)$. Fourth row: temperature. Time from left to right: $1 \mu$s, $3 \mu$s, and $6 \mu$s. \label{fig:Mach100-TimeEvolution}}
\end{center}
\end{figure}

The volume fraction, pressure, schlieren and temperature, for this test are shown in Figure  \ref{fig:Mach100-TimeEvolution}. The simulation requires the positivity limiting procedure to remain stable. Even with the limiter present the shocks remain sharply captured by the WENO5Z \cite{borges2008improved} scheme and the volume fraction field shows that the interface stays immiscible and does not contain spurious oscillations even when interacting with large temperature and pressure gradients. 

\subsection{Multi-phase multi-component tests} \label{sec:MultiphaseMulticomponentTests}
This Section contains three multi-phase multi-component tests to explore the full abilities of the proposed formulation. The first case is a newly proposed one-dimensional Shu-Osher problem. The second case is a multi-component phase constrained diffusion problem. The final test is a two-layer cylindrical Richtmyer-Meshkov implosion to show interactions with interfaces, shocks, and component mixing.

\subsubsection{Multi-phase multi-component Shu-Osher}

To test the presence of shocks interacting with a regularized multi-phase multi-component interface, a modified Shu-Osher problem is proposed. The materials used in this test are defined using the ideal gas equation of state with different $\gamma$ as, $\gamma_1 = 1.4$, $\gamma_2 = 1.3$, $\gamma_3 = 1.885$. Though the materials are all gaseous, this case is computationally considered a multi-phase multi-component test as the interface between $\gamma_1$ is enforced as immiscible with $\gamma_2$ and $\gamma_3$ using the interface regularization terms. The initial condition is shown in Table \ref{tab:MPMC_ShuOsherSetp}. The profile defining the interface between $\gamma_2$ and $\gamma_3$ for $x\geq 1$ can be found using the relations, $X_2 = \frac{1/(\gamma - 1) - 1/(\gamma_3 - 1)}{1/(\gamma_2 -1) - 1/(\gamma_3 - 1)}$, and $X_3 = 1.0 - X_2$.

\begin{table}[hbt!]
    \centering
    \begin{tabular}{l c c c c c c}
    \hline
     Location & Material &  $\rho$  & u  & P  & $\mu$  & $\gamma$
    \\[1mm]
    \hline
    $0\leq x < 1$ & $X_1$ & 3.857143 & 2.629369 & 10.3333 & $0.0$ & 1.4 
    \\[1mm]
    $1\leq x < 20.0$ & $X_2$ $\&$ $X_3$ & $1 + 0.2\sin\left({5(x-5)}\right)$ & 0.0 & 1.0 & $0.0$ & $1 + 1/\left(1.33 + 0.2\sin\left(5(x-5)\right)\right)$
    \\[1mm]
    \hline
    \\
    \end{tabular}
    \caption{Initial conditions for 1D multi-phase multi-component Shu Osher}
    \label{tab:MPMC_ShuOsherSetp}
\end{table}

\begin{figure}[hbt!]
\begin{center}
\includegraphics[width=\textwidth]{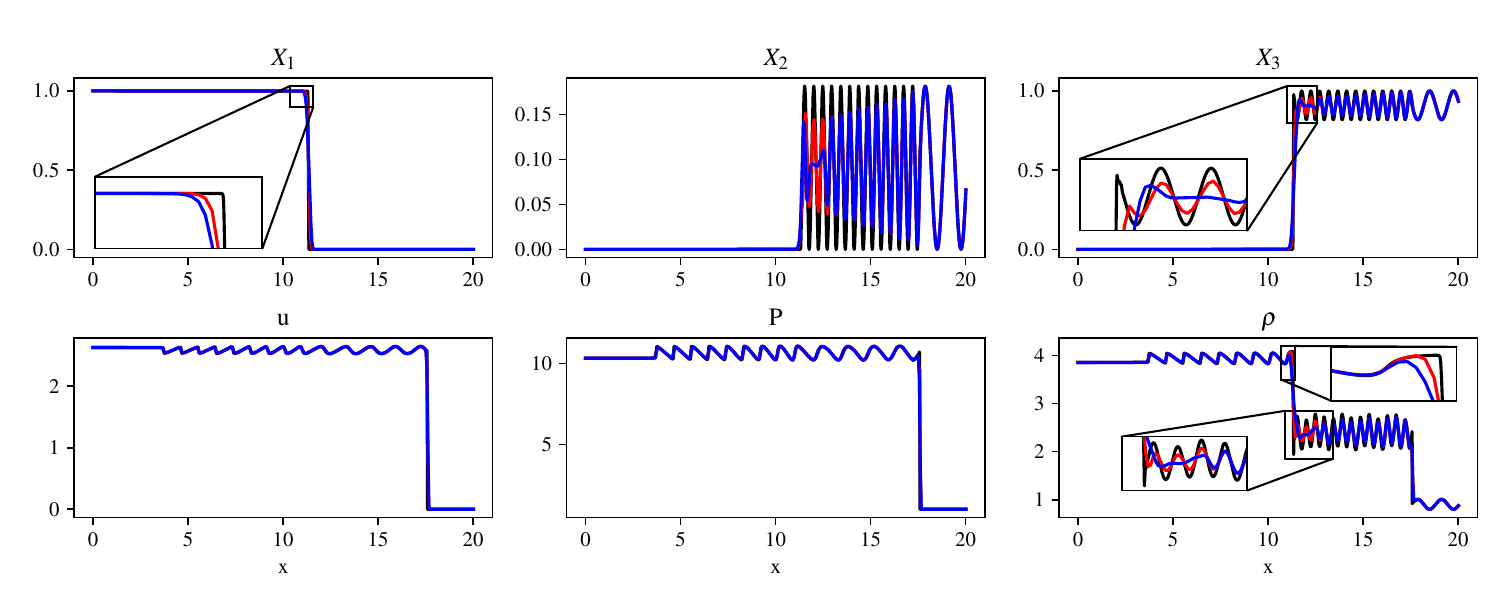}
\caption{Multi-phase multi-component modified Shu-Osher problem at time $t = 4$. Reference case (\protect\full{}), with CDI interface regularization (\protect\fullRed{}), and without interface regularization (implicit interface capturing) (\protect\fullBlue{}). \label{fig:MPMC-ShuOsher}}
\end{center}
\end{figure}

Figure \ref{fig:MPMC-ShuOsher} shows the results of the multi-phase multi-component Shu Osher simulation using a spatial resolution of $(400 \times 1)$ with and without interface regularization compared to a reference of spatial resolution $(12800 \times 1)$ without regularization at a time $t=4.0$. The regularization sharpens the interface between the immiscible phases and leads to a closer match with the reference compared to only using the ENO-type scheme. Additionally, the components within the gas phase individually behave as expected and do not contain spurious oscillation due to the introduction of the interphase regularization term. 

\subsubsection{Phase constrained multi-component mixing}

\begin{figure}[hbt!]
\begin{center}
\includegraphics[width=0.5\textwidth]{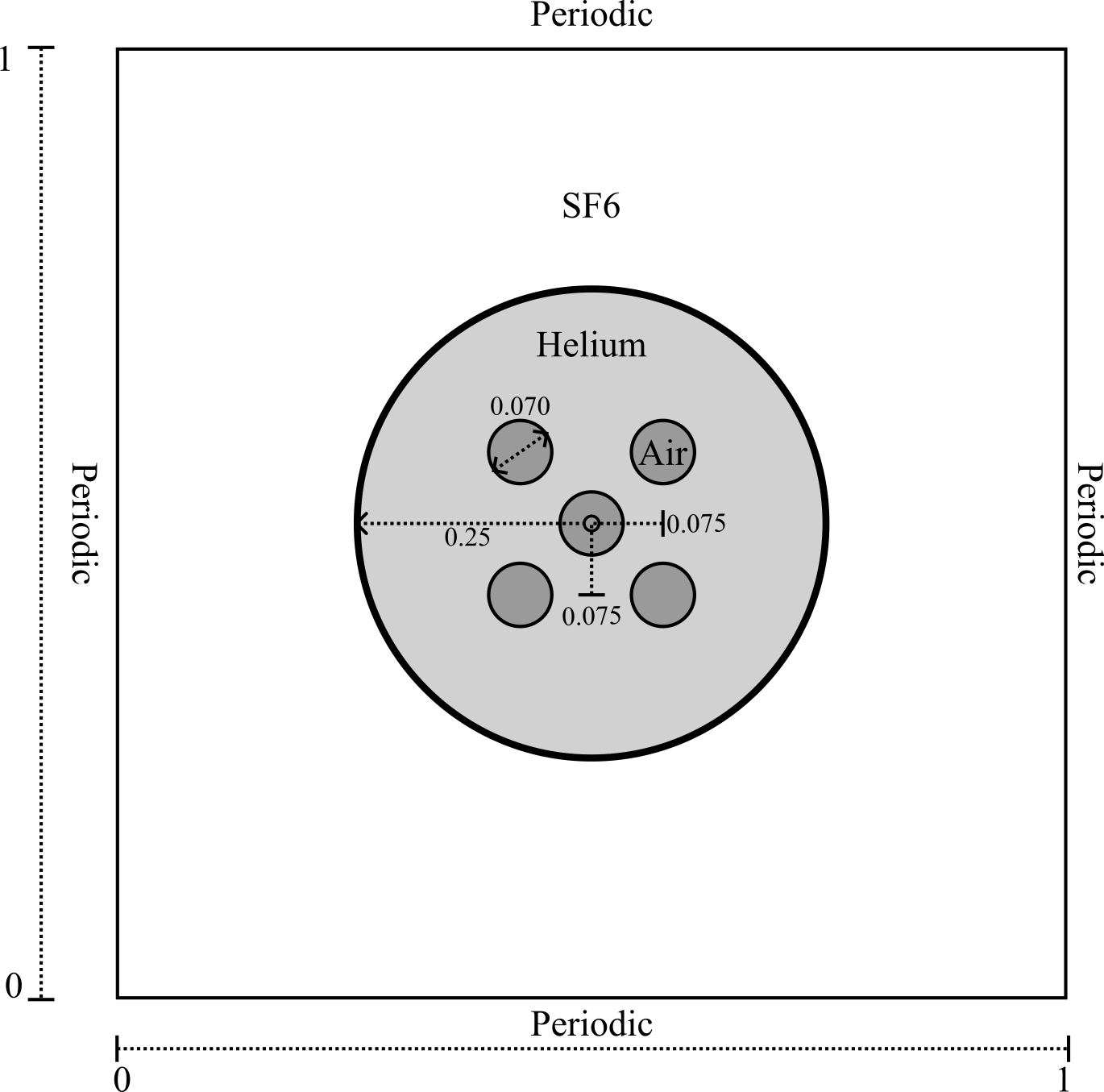}
\caption{Schematic of initial condition for phase constrained diffusion between air-He-SF6.\label{fig:PhaseConstrainedDiffusion}}
\end{center}
\end{figure}

The model proposed to capture intraphase diffusion without introducing leakage between phases was described in Eqs. \ref{eq:SpeciesMassDiffusion} and the final form is included here,

\begin{align}
  \label{eq:SpeciesMassDiffusionRepeated}
    J_{p,\bm x_i}^c & = - \rho  \left[ D^c_pY_p\frac{W_p^c}{W_p} \frac{\partial }{\partial \bm x_i}  \left(\frac{X_p^c}{X_p}\right) - Y_p^c\sum_j \left(\frac{W_p^j}{W_p}\right) D^j_p \frac{\partial }{\partial \bm x_i} \left(\frac{X_p^j}{X_p}\right)\right].
\end{align}
To test this model, we simulate the diffusion between air and helium within a phase-constrained bubble surrounded by SF6. The goal of this case is to showcase the ability of Eq. \ref{eq:SpeciesMassDiffusion} to allow for intraphase diffusion without leaking material across the immiscible interface boundary. To study the behavior of the model we propose a case which consists of an immiscible interface between SF6 and air/helium being enforced throughout time. Additionally, within the helium bubble, five air pockets are initialized to induce intraphase mixing. The initial condition for this case is shown in Figure \ref{fig:PhaseConstrainedDiffusion} and the spatial resolution is ($400 \times 400$). The thermodynamic state for this problem is defined in Table \ref{tab:PhaseConstrainedDiffustionSetup}, where the physical viscosity of all components was increased by a factor of 10,000 to speed up the simulation by ensuring that the time step is limited by diffusion. 

\begin{table}[hbt!]
    \centering
    \begin{tabular}{l c c c c c }
    \hline
     Material &  $\rho$ [kg/$\text{m}^3$] & u [m/s] & P [Pa] & $\mu$ [Pa-s] & Sc
    \\[1mm]
    \hline  
    SF6  & 6.06 & 0.0 & 101325.0 & $1.5 \times 10^{-1}$ & - \\[1mm]
    Air  & 1.18 & 0.0 & 101325.0 & $1.81 \times 10^{-1}$ & 1.0
    \\[1mm]
    Helium  & 0.166 & 0.0 & 101325.0 & $1.96 \times 10^{-1}$ & 0.7
    \\[1mm]
    \hline
    \end{tabular}
    \caption{Thermodynamic state for intraphase diffusion between air and helium surrounded by SF6. For this problem only, viscosity values are artificially increased by a factor of 10000 to speed up convergence to the equilibrium state by making the solution diffusion dominated.} \label{tab:PhaseConstrainedDiffustionSetup}
\end{table}

\begin{figure}[hbt!]
\begin{center}
\includegraphics[width=0.85\textwidth]{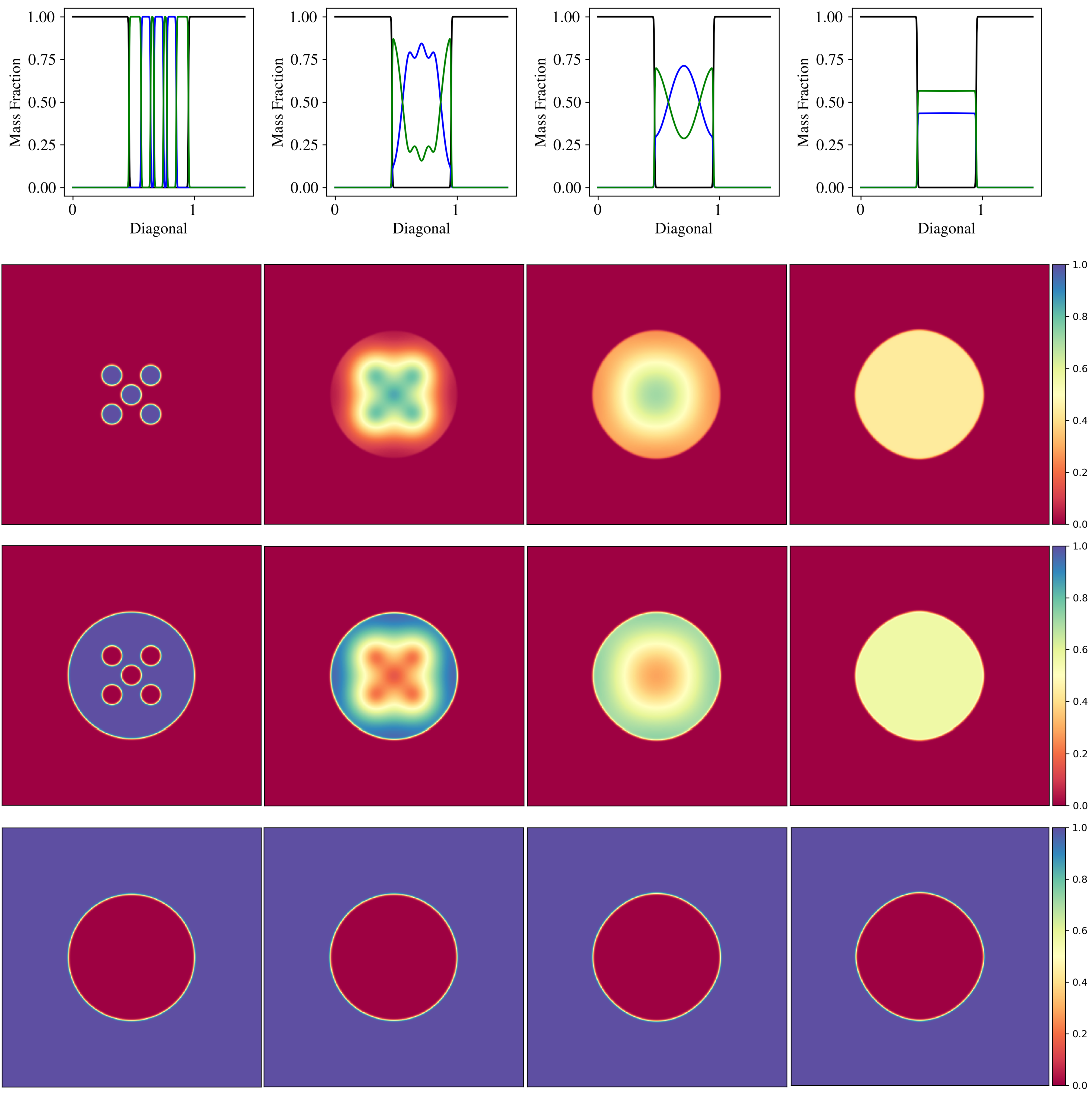}
\caption{Time evolution of phase constrained diffusion between air-He-SF6. First row: 1D line plot of mass fraction profiles for air (\protect\fullBlue{}), Helium (\protect\fullGreen{}), and SF6 (\protect\full{}) along the diagonal. Second row: air. Third row: Helium. Bottom row: SF6. Times from left to right: $0$ ms, $2$ ms, $6$ ms, $41.5$ ms. \label{fig:PhaseConstrainedDiffusion-TimeEvolution}}
\end{center}
\end{figure}

Figure  \ref{fig:PhaseConstrainedDiffusion-TimeEvolution} shows the evolution of air, helium, and SF6 throughout time. Throughout the diffusion process there is no leakage of air/helium into the SF6 and the interface remains immiscible. Additionally, the intraphase diffusion model results in the pockets of air mixing with the helium and eventually converging to a constant state. 

\subsubsection{Two-layer Richtmyer-Meshkov instability implosion}

\begin{figure}[hbt!]
\begin{center}
\includegraphics[width=0.5\textwidth]{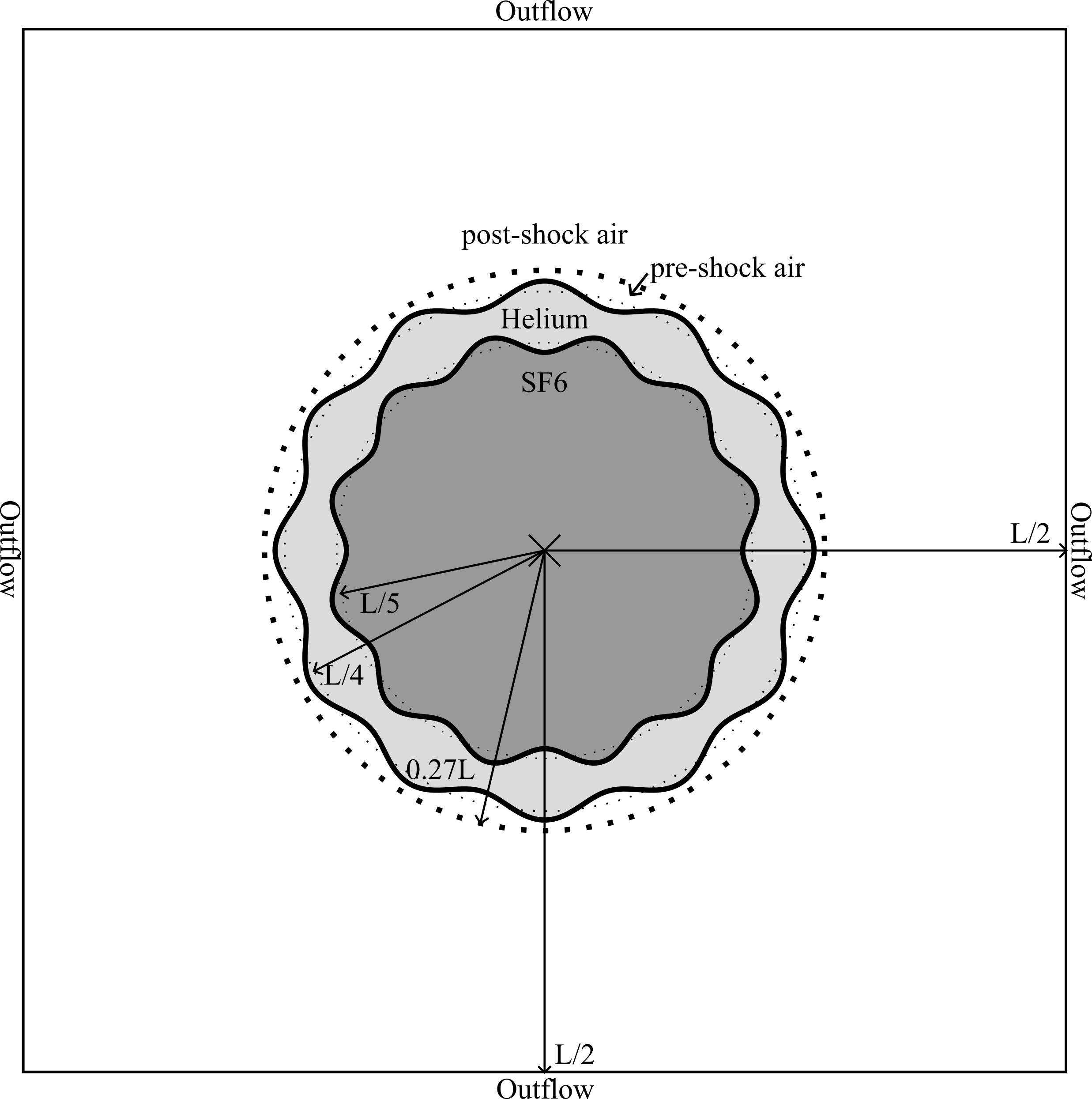}
\caption{Schematic of initial condition for two-layer Rightmyer-Meshkov implosion between air-He-SF6.\label{fig:TwoLayerRMISchematic}}
\end{center}
\end{figure}

\begin{figure}[hbt!]
\begin{center}
\includegraphics[width=\textwidth]{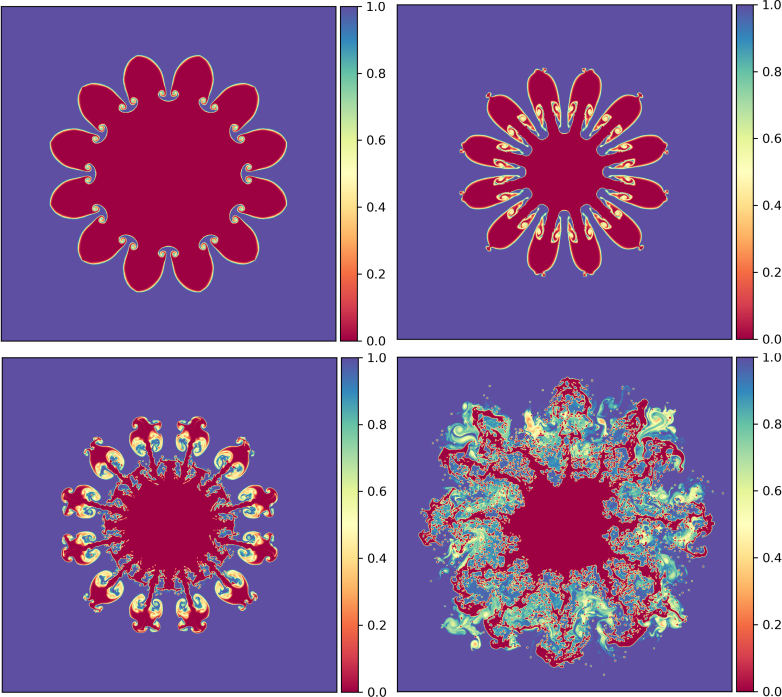}
\caption{Mass fraction of air during two-layer Rightmyer-Meshkov implosion between air-He-SF6 at times $11.6 \mu$s, $18.7 \mu$s, $27.1 \mu$s, and $47 \mu$s.\label{fig:TwoLayerRMI_Density}}
\end{center}
\end{figure}

\begin{figure}[hbt!]
\begin{center}
\includegraphics[width=\textwidth]{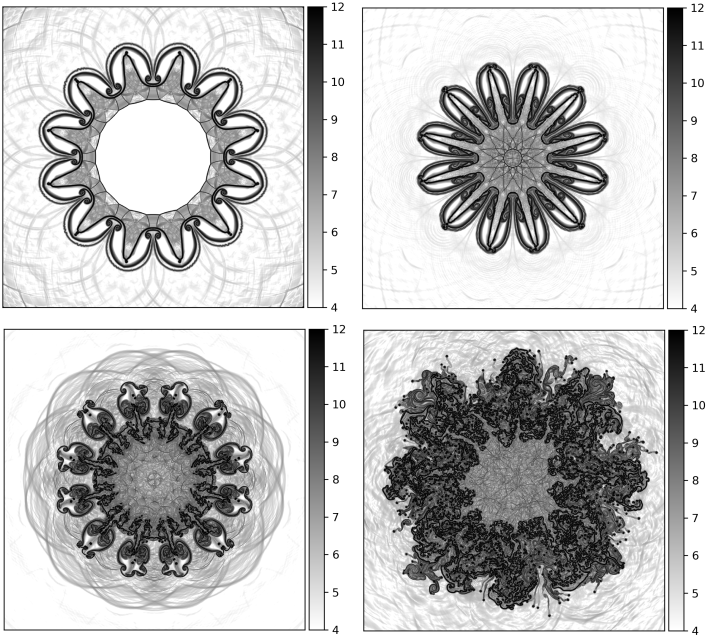}
\caption{Numerical Schlieren $\left(\ln\left(\frac{||\nabla \rho||}{\rho}\right)\right)$ results for two-layer Rightmyer-Meshkov implosion between air-He-SF6 at times $11.6 \mu$s, $18.7 \mu$s, $27.1 \mu$s, and $47 \mu$s..\label{fig:TwoLayerRMI_Schlieren}}
\end{center}
\end{figure}

This section contains the final multi-phase multi-component test of a two-layer Rightmyer-Meshkov implosion problem. This type of flow can be thought of as a model problem to represent many of the hydrodynamic physics present in Inertial Confinement Fusion (ICF). The initial condition for this case is shown in Figure  \ref{fig:TwoLayerRMISchematic}, where there are three materials, air, helium, and SF6, separated by perturbed material interfaces which will be simulated with a spatial resolution of ($3200 \times 3200$). The initial thermodynamic state is defined in Table \ref{tab:RMI2Layer-SF6AirHeliumSetup}. In this simulation the air and helium are treated as one phase where intraphase mixing occurs, and the air/helium-SF6 interface is treated as immiscible. A Mach 1.22 shock is initialized in air and will first cross the air-helium interface and induce a RMI which will lead to intraphase diffusion between the air and helium. Later, the shock passes through the interface between the air-helium phase with the SF6 phase creating another RMI where the immiscibility condition is enforced. At later times the shock implodes at the center of the domain and reverses into a outward traveling pressure wave which breaks the existing RMI interface structure into a chaotic field. Figures \ref{fig:TwoLayerRMI_Density} and \ref{fig:TwoLayerRMI_Schlieren} show the evolution of the mass-fraction of air and Schlieren, respectively. The evolution of the mass-fraction and Schlieren both show the regimes of the implosion problem, including the formation of the intraphase and interphase RMIs as well as the eventual transition to a chaotic state. In particular, Figure \ref{fig:TwoLayerRMI_Density} shows the mixing of air and helium through the simulation, whereas the interface between air and SF6 remains immiscible due to the multi-component extension of the CDI interface regularization model. The regimes of this simulation include interactions between shocks, immiscible interfaces, and multi-component mixing. Specifically, the shock remains stable and non-oscillatory by the ENO-type scheme, the SF6-air/helium interface remains immiscible due to the regularization terms, and the air and helium are mixed due to the phase-constrained diffusion model.

\begin{table}[hbt!]
    \centering
    \begin{tabular}{l c c c c c }
    \hline
     Material &  $\rho$ [kg/$\text{m}^3$] & u$_r$ [m/s] & P [Pa] & $\mu$ [Pa-s] & $Sc$ [-]
    \\[1mm]
    \hline
    SF6 & 6.06 & 0.0 & 101325.0 & $1.5\times10^{-5}$ & - \\[1mm]
    Helium  & 0.166 & 0.0 & 101325.0 & $1.96\times 10^{-5}$ & $0.70$
    \\[1mm]
    pre-shock air  & 1.18 & 0.0 & 101325.0 & $1.81\times 10^{-5}$ & $1.0$
    \\[1mm]
    post-shock air  & 1.6638 & 125.3 & 164957.0 & $1.81\times 10^{-5}$ & $1.0$
    \\[1mm]
    \hline
    \\
    \end{tabular}
    \caption{Initial conditions for two-layer RMI. The inward radial velocity is denoted u$_r$}.
    \label{tab:RMI2Layer-SF6AirHeliumSetup}
\end{table}

\section{Conclusions} \label{sec:Conculsion}

We have presented a robust and concise computational framework to simulate multi-phase multi-component flows. A positivity-preserving ENO-type scheme is consistently designed with the thermal-mechanical assumptions of the multi-phase model to achieve oscillation free solutions for shocks and material interfaces. Specifically, with the proposed method the interface equilibrium condition (IEC) was discretely satisfied without requiring additional of redundant PDEs. Additionally, the conservative diffuse interface (CDI) method was generalized to multi-phase multi-component systems by assuming intraphase mixing was occurring between scalar fields confined to immiscible phases. The extended CDI model allows for the simultaneous representation of immiscible interfaces and intraphase mixing. Similar ideas were used to confine the physical species diffusion to individual phases without leakage across immiscible interfaces. 

The proposed framework was used to simulate problems ranging in complexity, from single-phase to multi-phase multi-component flows. Single-phase tests included a 2D Riemann problem which showed the high-resolution capabilities of the proposed framework. Multi-component tests included a single-mode Richtmyer-Meshkov instability which verified the results of the proposed framework with those of former computational studies. The multi-phase tests showed the ability of the framework to handle the interaction of high-density immiscible phase interfaces with shocks. Furthermore, a simulation of a Mach 100 water column was completed without simulation failure, showing the usefulness of the positivity preserving procedure. Lastly, multiple multi-phase multi-component simulations showcased the interactions of the generalized CDI model, the intraphase species diffusion model, and the ENO-type shock capturing. In particular, a high-resolution ICF-like RMI implosion problem showed the ability of the framework to simulate complex multi-physics systems involving shocks, intraphase mixing, and interphase immiscibility. 

Future work includes incorporating phase change, surface tension, and reactions. Additionally, extensions of the equation of state to include non-linear equation of state dependence on temperature will be required to study combustion and other high-temperature related problems. The applicability of these additions will provide a path for high-resolution simulations of many engineering systems, including engines and rocket combustors.
\appendix

\section{Characteristic decomposition} \label{sec:CharacteristicForm}

The Euler equations (ignoring viscosity and interface regularization in Eq. \ref{eq:GoverningEquations}) is defined by,
\begin{align} \label{eq:EulerEquations}
    \frac{\partial \bm U}{\partial t} + \frac{\partial \left [\mathcal{F}(\bm U)\right ]}{\partial x} + \frac{\partial \left [\mathcal{G}(\bm U)\right ]}{\partial y} + \frac{\partial \left[\mathcal{H}(\bm U)\right ]}{\partial z} = 0.
\end{align}
where it can be re-written as in terms of primitive variables, $\bm W = [T, Y^c_p,u,v,w,P]$, using a Jacobian matrix of conserved and primitive variables defined as,
\begin{equation}
    \bm{P} = \frac{\partial \bm{U}}{\partial \bm{W}}
\end{equation}
and flux Jacobian matrices defined by,
\begin{equation}
    \bm{Q_x} = \frac{\partial \bm{\mathcal{F}}}{\partial \bm{W}}, \quad     \bm{Q_y} = \frac{\partial \bm{\mathcal{G}}}{\partial \bm{W}}, \quad     \bm{Q_z} = \frac{\partial \bm{\mathcal{H}}}{\partial \bm{W}}.
\end{equation}
Now the quasi-linearized Euler equation in terms of primitive variables can be written as,
\begin{align}
\label{eq:EulerEquationsPrimitiveLinearized}
    \frac{\partial \bm W}{\partial t} + \bm A\frac{\partial \bm W}{\partial x} + \bm B\frac{\partial \bm W}{\partial y} + \bm C\frac{\partial \bm W}{\partial z}= 0.
\end{align}
where $\bm{A} = \bm{P}^{-1}\bm{Q}_x$, $\bm{B} = \bm{P}^{-1}\bm{Q}_y$, and $\bm{C} = \bm{P}^{-1}\bm{Q}_z$ are defined as,
%
\begin{equation}
    \bm A = \begin{bmatrix} \setlength{\arraycolsep}{1pt}
        u & \zeroCdots & \frac{\rho a^2\beta-1}{\alpha} & 0 & 0 & 0 \\ 
        \zeroCvdots & \uCddots  & \zeroCvdots & \zeroCvdots & \zeroCvdots & \zeroCvdots \\
        0 & 0  & u & 0 & 0 & \frac{1}{\rho} \\
        0 & 0  & 0 & u & 0 & 0 \\
        0 & 0  & 0 & 0 & u & 0 \\
        0 & 0  & \rho a^2 & 0 & 0 & u
    \end{bmatrix}  
    \bm B = \begin{bmatrix} \setlength{\arraycolsep}{1pt}
        v & \zeroCdots & 0 &\frac{\rho a^2\beta-1}{\alpha} & 0 & 0 \\ 
        \zeroCvdots & \vCddots  & \zeroCvdots & \zeroCvdots & \zeroCvdots & \zeroCvdots \\
        0 & 0  & v & 0 & 0 & 0 \\
        0 & 0  & 0 & v & 0 & \frac{1}{\rho} \\
        0 & 0  & 0 & 0 & v & 0 \\
        0 & 0  & 0 & \rho a^2 & 0 & v
    \end{bmatrix}  
    \bm C = \begin{bmatrix} \setlength{\arraycolsep}{1pt}
        w & \zeroCdots & 0 & 0 &\frac{\rho a^2\beta-1}{\alpha} & 0 \\ 
        \zeroCvdots & \wCddots  & \zeroCvdots & \zeroCvdots & \zeroCvdots & \zeroCvdots \\
        0 & 0  & w & 0 & 0 & 0 \\
        0 & 0  & 0 & w & 0 & 0 \\
        0 & 0  & 0 & 0 & w & \frac{1}{\rho} \\
        0 & 0  & 0 & 0 & \rho a^2 & w
    \end{bmatrix}
\end{equation}
where the $(\cdots)$ represent the number of additional entries required to represent all but one $Y_p^c$ species in the system. The primitive system can be diagonalized using a characteristic decomposition,
\begin{equation*}
    \mathbf{A} = \mathbf{S_A}\mathbf{\Lambda}\mathbf{S_A^{-1}} \quad     \mathbf{B} = \mathbf{S_B}\mathbf{\Lambda}\mathbf{S_B^{-1}} \quad     \mathbf{C} = \mathbf{S_C}\mathbf{\Lambda}\mathbf{S_C^{-1}}
\end{equation*}
where, for $\bm A$, the right-eigenvectors $\mathbf{S_A}$ and left-eignevectors $\mathbf{S_A^{-1}}$ are defined as,
\begin{equation}
    \mathbf{S_A} = \begin{bmatrix}
        1 & \zeroCdots &  0 & 0 & \frac{(\rho a^2\beta - 1)}{(\alpha a^2\rho)} & \frac{(\rho a^2\beta - 1)}{(\alpha a^2\rho)}  \\ 
        \zeroCvdots & \oneCddots  & \zeroCvdots & \zeroCvdots & \zeroCvdots & \zeroCvdots \\
        0 & 0  & 0 & 0 & 1/(a\rho) & -1/(a\rho) \\
        0 & 0  & 1 & 0 & 0 & 0 \\
        0 & 0  & 0 & 1 & 0 & 0 \\
        0 & 0  & 0 & 0 & 1 & 1
    \end{bmatrix}, \quad 
    \mathbf{S_A^{-1}} = \begin{bmatrix}
        1 & \zeroCdots &  0 & 0 & 0 & -\frac{(\rho a^2\beta - 1)}{(\alpha a^2\rho)}  \\ 
        \zeroCvdots & \oneCddots & \zeroCvdots & \zeroCvdots & \zeroCvdots & \zeroCvdots \\
        0 & 0  & 0 & 1 & 0 & 0 \\
        0 & 0  & 0 & 0 & 1 & 0 \\
        0 & 0  & (a\rho)/2 & 0 & 0 & 1/2 \\
        0 & 0  & -(a\rho)/2 & 0 & 0 & 1/2
    \end{bmatrix}.
\end{equation}

Similar characteristic decomposition can be performed for $\bm B$ as,
\begin{equation}
    \mathbf{S_B} = \begin{bmatrix}
        1 & \zeroCdots &  0 & 0 &\frac{(\rho a^2\beta - 1)}{(\alpha a^2\rho)} & \frac{(\rho a^2\beta - 1)}{(\alpha a^2\rho)}  \\ 
        \zeroCvdots & \oneCddots  & \zeroCvdots & \zeroCvdots & \zeroCvdots & \zeroCvdots \\
        0 & 0  & 1 & 0 & 0 & 0 \\
        0 & 0  & 0 & 0 & 1/(a\rho) & -1/(a\rho) \\
        0 & 0  & 0 & 1 & 0 & 0 \\
        0 & 0  & 0 & 0 & 1 & 1
    \end{bmatrix}, \quad     
    \mathbf{S_B^{-1}} = \begin{bmatrix}
        1 & \zeroCdots &  0 & 0 & 0 & -\frac{(\rho a^2\beta - 1)}{(\alpha a^2\rho)}  \\ 
        \zeroCvdots & \oneCddots & \zeroCvdots & \zeroCvdots & \zeroCvdots & \zeroCvdots \\
        0 & 0  & 1 & 0 & 0 & 0 \\
        0 & 0  & 0 & 0 & 1 & 0 \\
        0 & 0  & 0 & (a\rho)/2 & 0 & 1/2 \\
        0 & 0  & 0 & -(a\rho)/2 & 0 & 1/2
    \end{bmatrix}
\end{equation}
and additionally for $\bm C$, 
\begin{equation}
    \mathbf{S_C} = \begin{bmatrix}
        1 & \zeroCdots &  0 & 0 & \frac{(\rho a^2\beta - 1)}{(\alpha a^2\rho)} & \frac{(\rho a^2\beta - 1)}{(\alpha a^2\rho)}  \\ 
        \zeroCvdots & \oneCddots  & \zeroCvdots & \zeroCvdots & \zeroCvdots & \zeroCvdots \\
        0 & 0  & 1 & 0 & 0 & 0 \\
        0 & 0  & 0 & 1 & 0 &  0 \\
        0 & 0  & 0 & 0 & 1/(a\rho) & -1/(a\rho) \\
        0 & 0  & 0 & 0 & 1 & 1
    \end{bmatrix}, \quad 
    \mathbf{S_C^{-1}} = \begin{bmatrix}
        1 & \zeroCdots &  0 & 0 & 0 & -\frac{(\rho a^2\beta - 1)}{(\alpha a^2\rho)}  \\ 
        \zeroCvdots & \oneCddots & \zeroCvdots & \zeroCvdots & \zeroCvdots & \zeroCvdots \\
        0 & 0  & 1 & 0 & 0 & 0 \\
        0 & 0  & 0 & 1 & 0 & 0 \\
        0 & 0  & 0 & 0 & (a\rho)/2 & 1/2 \\
        0 & 0  & 0 & 0 & -(a\rho)/2 & 1/2
    \end{bmatrix}.
\end{equation}
With the definitions for the characteristic composition the projections to/from characteristic space in the Godunov algorithm described in Section \ref{sec:GodunovApproach} can be completed. 

\section{HLLC Riemann solver} \label{sec:HLLC}

To solve the Riemann problem at the cell-faces we use the approximate 1D HLLC Riemann solver. Extension to multi-dimensional flows is done using a dimension-by-dimension approach. As an example, the following section will define the 1D HLLC approximate Riemann solver in the x-direction. The wave speeds are defined as,
\begin{align}
    s^L &= \min(\Bar{u}-\Bar{a}, u^L - a^L) \\
    s^R &= \max(\Bar{u}+\Bar{a}, u^R + a^R),
\end{align}
and
\begin{align}
    s^- &= \min(0, s^L) \\
    s^+ &= \max(0, s^R)
\end{align}
where the overbar represents the arithmetic average across the $x_{i+1/2}$ interface.

The intermediate wave speed is chosen as,

\begin{equation}
    s^* = \frac{P^R - P^L + \rho^Lu^L(s^L-u^L) - \rho^Ru^R(s^R - u^R)}{\rho^L(s^L - u^L) - \rho^R(s^R - u^R)}
\end{equation}
Lastly, the intermediate conservative state is defined by,
\begin{align} 
    \bm U^{*K} = \left(\frac{s^K - u^K}{s^K-s^*}\right)\begin{bmatrix}
    (\rho Y_{p}^c)^K \\
    \rho^Ks^* \\ \rho^Kv^K \\ \rho ^Kw^K \\ E^K + (s^* - u^K)\left(\rho^Ks^* + \frac{P^K}{s^K-u^K}\right)
    \end{bmatrix} 
\end{align}
With the wave speeds and the intermediate state defined the local Riemann problem at cell face $x_{i+1/2}$ can be solved using the HLLC Riemann solver to determine the convective flux, $\mathcal{F}_{i+1/2} = \bm {HLLC}(\bm U^L_{i+1/2}, \bm U^R_{i+1/2})$ where the $\bm {HLLC}$ flux can be defined symmetrically as,

\begin{equation}
    \bm {HLLC}(\bm U^L, \bm U^R) = \frac{1 + \text{sign}(s^*)}{2}\left(\mathcal{F}^L + s^-(\bm U^{*L} - \bm U^{L})\right) + \frac{1 - \text{sign}(s^*)}{2}\left(\mathcal{F}^R + s^+(\bm U^{*R} - \bm U^{R})\right).
\end{equation}

\section{IEC analysis with characteristic variables} \label{sec:IECCharacteristicAnalysis}
As discussed in Section \ref{sec:IEC} the clearest way to satisfy IEC is by directly interpolating the set of primitive variables $\bm W = [T, Y^c_p,u,v,w,P]$. Instead, in Section \ref{sec:DropAdvection} the characteristic variables projected from the primitive variables $\bm W$ is shown to provide acceptable levels of IEC error (near machine precision for multiple WENO-type schemes tested). The analysis showing how interpolations based on the characteristic variables remains IEC is shown below.

As an example, the characteristic variables in the x-direction based on $\bm W = [T, Y^c_p,u,v,w,P]$ is defined by, 
\begin{equation}
    \tilde{\bm W} = \mathbf{S_A^{-1}} \bm W = \begin{bmatrix}
        \tilde{w}_1 \\ 
        \tilde{w}_2 \\
        \tilde{w}_3 \\
        \tilde{w}_4 \\
        \tilde{w}_5 \\
        \tilde{w}_6
    \end{bmatrix} = \begin{bmatrix}
        T  - \overline{\frac{(\rho a^2\beta - 1)}{(\alpha a^2\rho)}} P \\ 
        Y^c_p \\
        v \\
        w \\
        u\overline{(a\rho)/2} + 1/2P \\
        -u\overline{(a\rho)/2} + 1/2P
    \end{bmatrix}.
\end{equation}
where the $(\overline{\cdot})$ represents the arithmetic (or Roe) average onto the face. These quantities are constant across the full stencil. Following the Godunov approach described in Section \ref{sec:GodunovApproach} $\tilde {\bm W}$ is projected using an ENO-type scheme to the left and right states across a cell-face. $\tilde{\bm W}^{L/R}$ are projected to $\bm W^{L/R}$ using,
\begin{equation}
    \bm W^{L/R} = \mathbf{S_A} \tilde{\bm W}^{L/R}
\end{equation}
To satisfy IEC after interpolation the following conditions must be met (assuming an initially constant $T, P, $ and $ \bm u$),
\begin{equation}
    T = T^{L/R}, \quad P = P^{L/R},  \quad \bm u = \bm u^{L/R}.
\end{equation}
First, the pressure and normal velocity conditions can be analyzed,
\begin{align}
    P^{L/R} &= \tilde{w}_5^{L/R} + \tilde{w}_6^{L/R} \\
    u^{L/R} &= \overline{(1/a\rho)}\tilde{w}_5^{L/R} - \overline{(1/a\rho)}\tilde{w}_6^{L/R}.
\end{align}
To show that pressure and velocity stay constant after ENO-type interpolation, we can note that both $\tilde{w}_5$ and $\tilde{w}_6$ are constant across an isothermal material interface by seeing that pressure, velocity, and the averaged quantity, $\overline{(a\rho)/2}$, are constant over the stencil. With both $\tilde{w}_5^{L/R} = \tilde{w}_5$ and $\tilde{w}_6^{L/R} = \tilde{w}_6$ we can see that $P^{L/R} = P$ and $u^{L/R} = u$.

The temperature on the cell-face is determined by,
\begin{align}
    T^{L/R} &= \tilde{w}_1^{L/R} + \overline{\frac{(\rho a^2\beta - 1)}{(\alpha a^2\rho)}}(\tilde{w}_5^{L/R} + \tilde{w}_6^{L/R}) \\
    &= \tilde{w}_1^{L/R} + \overline{\frac{(\rho a^2\beta - 1)}{(\alpha a^2\rho)}}(P^{L/R}) \\
    &= \left(T  - \overline{\frac{(\rho a^2\beta - 1)}{(\alpha a^2\rho)}} P\right)^{L/R} + \overline{\frac{(\rho a^2\beta - 1)}{(\alpha a^2\rho)}}(P^{L/R})
\end{align}
To see how temperature equilibrium is maintained, note that $\tilde{w}_1$ is also constant (across an isothermal interface), so $\tilde{w}_1^{(L/R)} = \tilde{w}_1$ leading to,
\begin{align}
    T^{L/R} &= T  - \overline{\frac{(\rho a^2\beta - 1)}{(\alpha a^2\rho)}} P + \overline{\frac{(\rho a^2\beta - 1)}{(\alpha a^2\rho)}}(P^{L/R})
\end{align}
where if the condition for $P = P^{L/R}$ is satisfied, $T^{L/R} = T$. 

The analysis above proves that using the characteristic variables based on the primitive basis vector $\bm W$ satisfies IEC by keeping pressure, temperature, and velocity oscillation free across an isothermal material interface. The theoretical analysis was confirmed in practice as the results of the IEC test in Section \ref{sec:DropAdvection} provided levels of error near machine precision for all fields.

\section{Single phase tests} \label{sec:SinglePhaseTests}

This Section contains three single-phase verification cases. The first is a two-dimensional vortex advection problem to verify the convergence of the numerical scheme in multiple dimensions. The second test is the standard Shu-Osher shock tube problem. The third is a common 2D Riemann problem to show the high-resolution capabilities of the Godunov implementation described in Section \ref{sec:GodunovApproach}.

\subsection{Two-dimensional vortex advection}

\begin{figure}[hbt!]
\begin{center}
\includegraphics[width=0.5\textwidth]{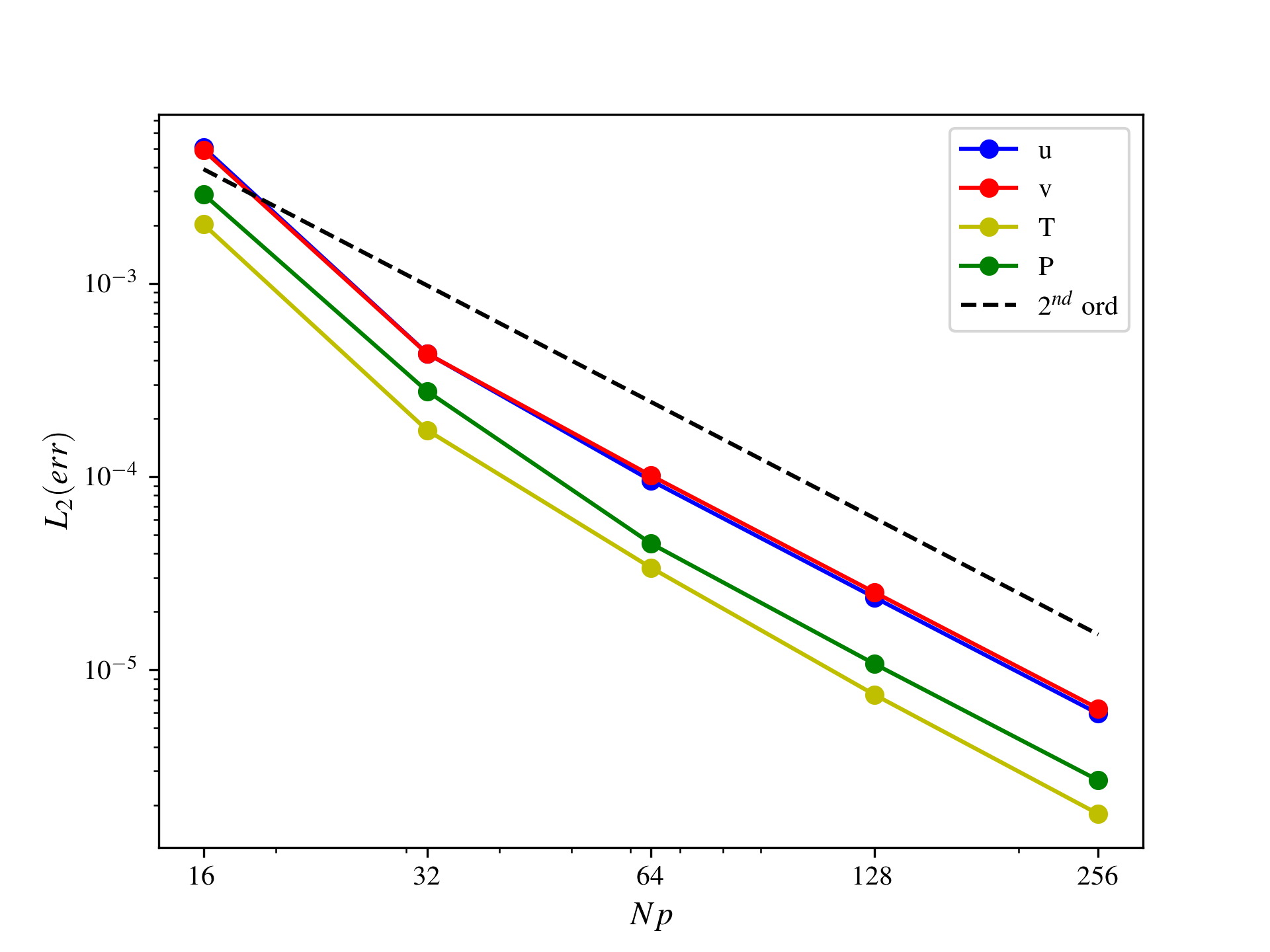}
\caption{Convergence of WENO5Z scheme for advection of 2D vortex. \label{fig:VortexAdvection}}
\end{center}
\end{figure}

To verify the implementation of the numerical scheme, a 2D vortex advection test was completed. Details on the initial condition for the test can be found in the following reference \cite{DiRenzo2020}. As mentioned in section \ref{sec:spatialDiscretization} a 1st-order quadrature rule was used to construct the cell-face averages. This common approach keeps the Godunov scheme at a reasonable cost, but reduces the order of accuracy to 2nd-order in multiple dimensions. Even so, the high-order ENO-type interpolations result in a low-dissipation and high-resolution scheme which, in practice, resembles the truly high-order counterpart at a far reduced cost. Figure \ref{fig:VortexAdvection} shows the expected order of convergence for the vortex advection problem using the WENO5Z \cite{borges2008improved} scheme with the Godunov method.

\subsection{Shu-Osher shock tube}

\begin{figure}[hbt!]
\begin{center}
\includegraphics[width=0.75\textwidth]{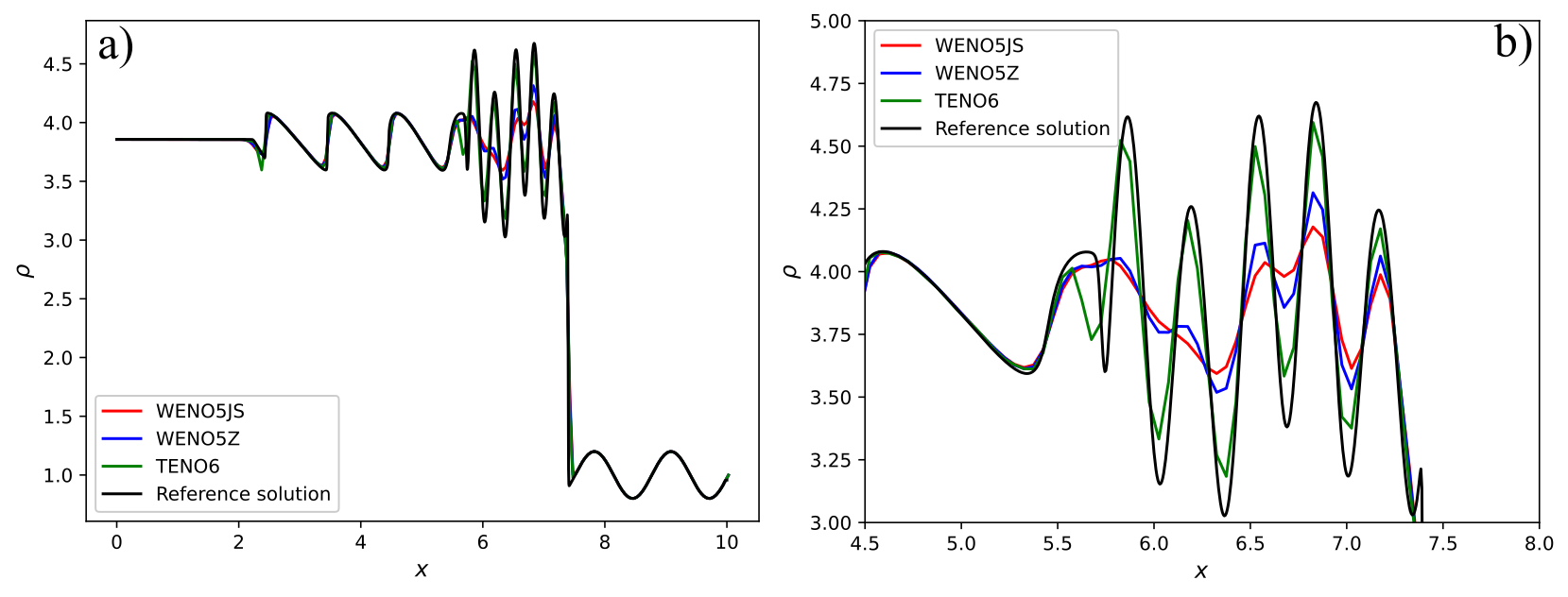}
\caption{Shu-Osher problem for (a) density and (b) density zoomed. Scheme comparisons for WENO5JS, WENO5Z, and TENO6 using Godunov scheme.\label{fig:ShuOsher}}
\end{center}
\end{figure}

Another standard single-phase test to show the applicability of the proposed scheme to handle shocks in single-phase flow is the Shu-Osher problem. The initial condition is given in Table \ref{tab:ShuOsherSetp} with a spatial resolution of $(200 \times 1)$ and a constant time step of $4\times 10^{-3}$. The resulting density field is shown at $t = 1.8$ in Figure \ref{fig:ShuOsher} to compare the influence of WENO5JS \cite{jiang1996efficient}, WENO5Z \cite{borges2008improved}, and TENO6 \cite{fu2016family} on the solution. As expected, the lowest dissipation scheme is TENO6 with a very close match to the high-resolution reference solution. Additionally, the improvements associated with the WENO5Z compared to the WENO5JS scheme are visually confirmed. 

\begin{table}[hbt!]
    \centering
    \begin{tabular}{l c c c c c}
    \hline
     Location &  $\rho$  & u  & P  & $\mu$  & $\gamma$
    \\[1mm]
    \hline
    $0\leq x < 1$  & 3.857143 & 2.629369 & 10.3333 & $0.0$ & 1.4 
    \\[1mm]
    $1\leq x < 10.0$ & $1 + 0.2\sin\left({5(x-5)}\right)$ & 0.0 & 1.0 & $0.0$ & 1.4
    \\[1mm]
    \hline
    \\
    \end{tabular}
    \caption{Initial conditions for 1D multi-phase multi-component Shu Osher}
    \label{tab:ShuOsherSetp}
\end{table}

\subsection{2D Riemann problem}

\begin{figure}[hbt!]
\begin{center}
\includegraphics[width=0.75\textwidth]{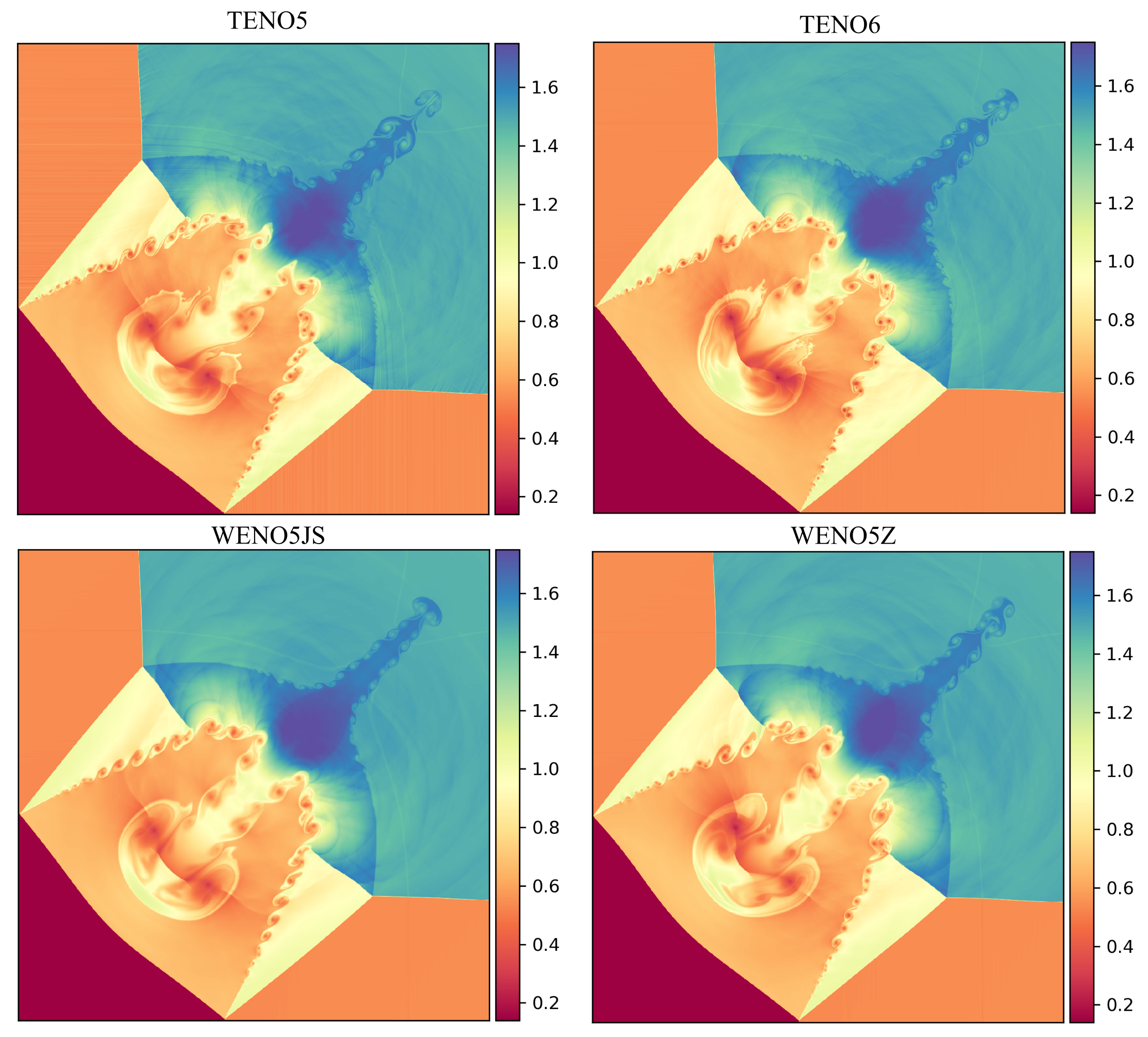}
\caption{Riemann problem comparing density fields using the high-resolution Godunov method with TENO5 \cite{fu2016family}, TENO6 \cite{fu2016family}, WENO5JS \cite{jiang1996efficient}, and WENO5Z \cite{borges2008improved} schemes. .\label{fig:Riemann2D}}
\end{center}
\end{figure}

The final single-phase test is a two-dimensional Riemann problem \cite{schulz1993numerical,fleischmann2019numerical} which will showcase the high-resolution capability of the proposed scheme. The initial conditions for the simulation are listed in Table \ref{tab:Riemann2D} for a domain of size $[0,2] \times [0, 2]$ with a spatial resolution of $(2000 \times 2000)$. Figure \ref{fig:Riemann2D} compares the density field for four schemes at a non-dimensional time $t=1.1$ over a subset of the domain defined by $[0,1.2]\times[0,1.2]$. The test case is without physical viscosity ($Re=\infty$), so small-scale structures are only destroyed by numerical dissipation. Figure \ref{fig:Riemann2D} shows the expected trends of more small-scale features for the lower dissipation TENO-type schemes compared to the WENO-type. Even so, all schemes showcase high-resolution solutions with both oscillation-free shocks and a richness of fine-scale structures. 

\begin{table}[hbt!]
    \centering
    \begin{tabular}{l c c c c c c}
    \hline
     Location &  $\rho$  & u  & v  & P  & $\mu$  & $\gamma$
    \\[1mm]
    \hline
    $x > 1$ \& $y > 1$  & 1.5 & 0.0 & 0.0 & 1.5 & 0.0 & 1.4 
    \\[1mm]
    $x < 1$ \& $y > 1$  & 33/62 & $4/\sqrt{11}$ & 0.0 & 0.3 & 0.0 & 1.4 
    \\[1mm]
    $x < 1$ \& $y < 1$  & 77/558 & $4/\sqrt{11}$ & $4/\sqrt{11}$ & 9/310 & 0.0 & 1.4 
    \\[1mm]
    $x > 1$ \& $y < 1$  & 33/62 & 0.0 & $4/\sqrt{11}$ & 0.3 & 0.0 & 1.4 
    \\[1mm]
    \hline
    \\
    \end{tabular}
    \caption{Initial conditions for the 2D Riemann problem}
    \label{tab:Riemann2D}
\end{table}

\section{Mixing rule definitions and identities} \label{sec:Mixing rules}
Conversions between mass-fraction, density, and volume fraction:
\begin{enumerate}
    \item  $\sum_p \sum_c Y_p^c = 1$
    \item  $\sum_c Y_p^c = Y_p$. Recall, $c$ represents the components of phase $p$. 
    \item $\rho_p \phi_p = \rho Y_p$
    \item $\sum_p\sum_c \rho Y_p^c = \sum_p \rho Y_p = \sum_p \phi_p \rho_p = \rho$    
    \item $\rho Y_p^c = \rho_p \phi_p Y_p^c/Y_p$
\end{enumerate}
Relations between mass-fraction, molar-fraction, and molecular weight:
\begin{enumerate}
    \item $W = \sum_p\sum_c X_p^c W_p^c = \left(\sum_p\sum_c Y_p^c/W_p^c\right)^{-1}$
    \item $W_p = \sum_c (X_p^c/X_p)W_p^c = \left(\sum_c (Y_p^c/Y_p)/W_p^c\right)^{-1}$
    \item $X_c^p = Y_c^pW/W_p^c$
\end{enumerate}

\bibliographystyle{elsarticle-num}
\bibliography{references}

\end{document}

%% file: Macro.tex
\newcommand{\comment}[1]{}

